\documentclass[pdflatex,sn-mathphys-ay]{sn-jnl}

\usepackage{graphicx}%
\usepackage{multirow}%
\usepackage{amsmath,amssymb,amsfonts}%
\usepackage{amsthm}%
\usepackage{mathrsfs}%
\usepackage[title]{appendix}%
\usepackage{xcolor}%
\usepackage{textcomp}%
\usepackage{manyfoot}%
\usepackage{booktabs}%
\usepackage{array}
\usepackage{algorithm}%
\usepackage{algorithmicx}%
\usepackage{algpseudocode}%
\usepackage{listings}%
\usepackage{multirow}
\usepackage{booktabs}
\usepackage{enumitem}
\usepackage{array}
\usepackage{makecell}
\usepackage{xurl}

\setcitestyle{authoryear,open={(},close={)},citesep={; }}

\theoremstyle{thmstyleone}%
\theoremstyle{thmstyletwo}%

\theoremstyle{thmstylethree}%

\begin{document}

\title[Article Title]{Revisiting SSL for sound event detection: complementary fusion and adaptive post-processing}


\author[1]{\fnm{Hanfang} \sur{Cui}}\email{1000530599@smail.shnu.edu.cn}

\author[1]{\fnm{Longfei} \sur{Song}}\email{songlongfei2@163.com}

\author[1]{\fnm{Li} \sur{Li}}\email{lili\_a0@163.com}

\author[2]{\fnm{Dongxing} \sur{Xu}}\email{xudongxing@unisound.com}

\author[1,2]{\fnm{Yanhua} \sur{Long}}\email{yanhua@shnu.edu.cn}

\affil[1]{
Shanghai Normal University, Shanghai, China}
\affil[2]{
Unisound AI Technology Co., Ltd., Beijing, China }


\abstract{

Self-supervised learning (SSL) models offer powerful representations for sound event detection (SED), 
yet their synergistic potential remains underexplored. This study systematically evaluates state-of-the-art 
SSL models to guide optimal model selection and integration for SED. We propose a framework combining 
heterogeneous SSL model (BEATs, HuBERT, WavLM, etc.) representations through three fusion strategies: 
individual SSL embedding integration, dual-modal fusion, and full aggregation. Our experiments on the DCASE 
2023 Task 4 Challenge reveal that dual-modal fusion (e.g., CRNN+BEATs+WavLM) achieves complementary performance gains, 
while CRNN+BEATs alone attains superior individual SSL embedding integration results. 
We further introduce a normalized sound event bounding boxes (nSEBBs), an adaptive post-processing method 
that dynamically adjusts target events detection boundaries, improving PSDS$_1$ by up to 4\% for 
standalone SSL models. These findings provide important insights into SSL model compatibility, 
demonstrating that task-specific fusion and dynamic post-processing enhance robustness. 
Our work establishes a reference for selecting and integrating SSL architectures in SED systems, 
balancing efficiency and accuracy for real-world deployment. 

}




\maketitle

\section{Introduction}\label{sec1}
Sound is a fundamental element of daily life, conveying rich environmental information and playing a vital role in how humans perceive and interact with the world. Despite its ubiquity, machines still struggle to accurately recognize sound events, especially in complex and noisy acoustic environments. Sound Event Detection (SED), a core task in computational audio analysis, aims to identify and localize specific events by detecting their temporal boundaries and assigning semantic labels. However, the development of effective SED systems is often hindered by the high cost of acquiring strongly labeled data.

To alleviate this challenge, the research community introduced the Detection and Classification of Acoustic Scenes and Events (DCASE) challenge in 2013 \citep{khelwal2024sound}. Initially focused on polyphonic sound detection, DCASE has since evolved to incorporate practical constraints, including real-world deployability. For instance, the 2023 edition introduced subtasks such as computational efficiency evaluation and soft-label training. These developments have expanded the scope of SED, enabling applications in diverse domains: in smart environments, SED supports applications like infant distress detection and urban soundscape monitoring \citep{bello2018sound, bello2019sonyc, debes2016monitoring}. In industrial settings, it facilitates machine anomaly detection for predictive maintenance \citep{dohi2023description, lv2023unsupervised}. Meanwhile, healthcare systems apply SED for patient monitoring \citep{liu2024icsd, wang2024cross}, and multimedia platforms integrate it into automated content analysis pipelines \citep{ashraf2015audio}.

Over the years, a wide range of modeling techniques have been proposed for SED. Early approaches used Hidden Markov Models (HMMs) and Gaussian Mixture Models (GMMs) to represent the temporal and acoustic characteristics of sound events \citep{clarkson1998auditory}. These were later superseded by deep learning-based methods, including Convolutional Neural Networks (CNNs) \citep{lecun1998gradient}, Convolutional Recurrent Neural Networks (CRNNs) \citep{shi2016end}, and transformer-based architectures \citep{vaswani2017attention}, which better capture complex patterns and long-range dependencies. However, these approaches typically require large amounts of labeled data, and their performance tends to degrade in unseen environments or with limited supervision.

Recently, self-supervised learning (SSL) has gained attraction across domains by leveraging unlabeled data for representation learning \citep{oord2018representation}. In SED, SSL models are typically used as feature extractors and integrated into CRNN frameworks \citep{zheng2021effective}. This strategy effectively reduces the need for labeled data while improving model performance by capitalizing on learned representations from large audio corpora. Nevertheless, a key research gap remains: existing studies typically rely on a single SSL model, without exploring the comparative or complementary advantages of different architectures. For example, CNN-based encoders are well-suited for capturing local spectral patterns, while transformer-based models 
demonstrate strong capabilities in modeling global temporal context. Despite this, the synergistic potential of combining multiple SSL models for SED remains underexplored.

In this work, we bridge this gap by proposing a novel SED framework that integrates multiple pre-trained SSL models and introduces a new post-processing method to enhance performance. Our main contributions are as follows:
\begin{itemize}[itemsep=10pt]

\item \textbf{Systematic Exploration of SSL Models for SED}: We benchmark several state-of-the-art SSL models on standard SED tasks and investigate multi-model fusion strategies to leverage complementary feature representations and improve overall detection accuracy. 

\item \textbf{Normalized SEBBs Post-processing}: We propose a novel normalization-based post-processing method, termed nSEBBs, which extends the original sound event bounding boxes (SEBBs) approach \citep{ebbers2024sound} 
to improve the system performance. We also perform a detailed comparison of single-threshold and double-threshold strategies to assess their respective impacts on SED performance. 

\end{itemize}

\section{Review of Related Works}
\label{sec2}

With the launch of DCASE Challenges, many studies have been proposed to enhance
the overall performance of SED systems. However, most of the recent works still focus on the 
three main areas, including:  a) developing more efficient CRNN-based architectures; b) 
exploring the use of different SSL representations as model inputs; and c) designing effective 
post-processing techniques to boost the detection performance.

\subsection{Sound Event Detection}
\label{subsec21}

\begin{figure}[ht]
    \centering
    \includegraphics[width=0.85\linewidth]{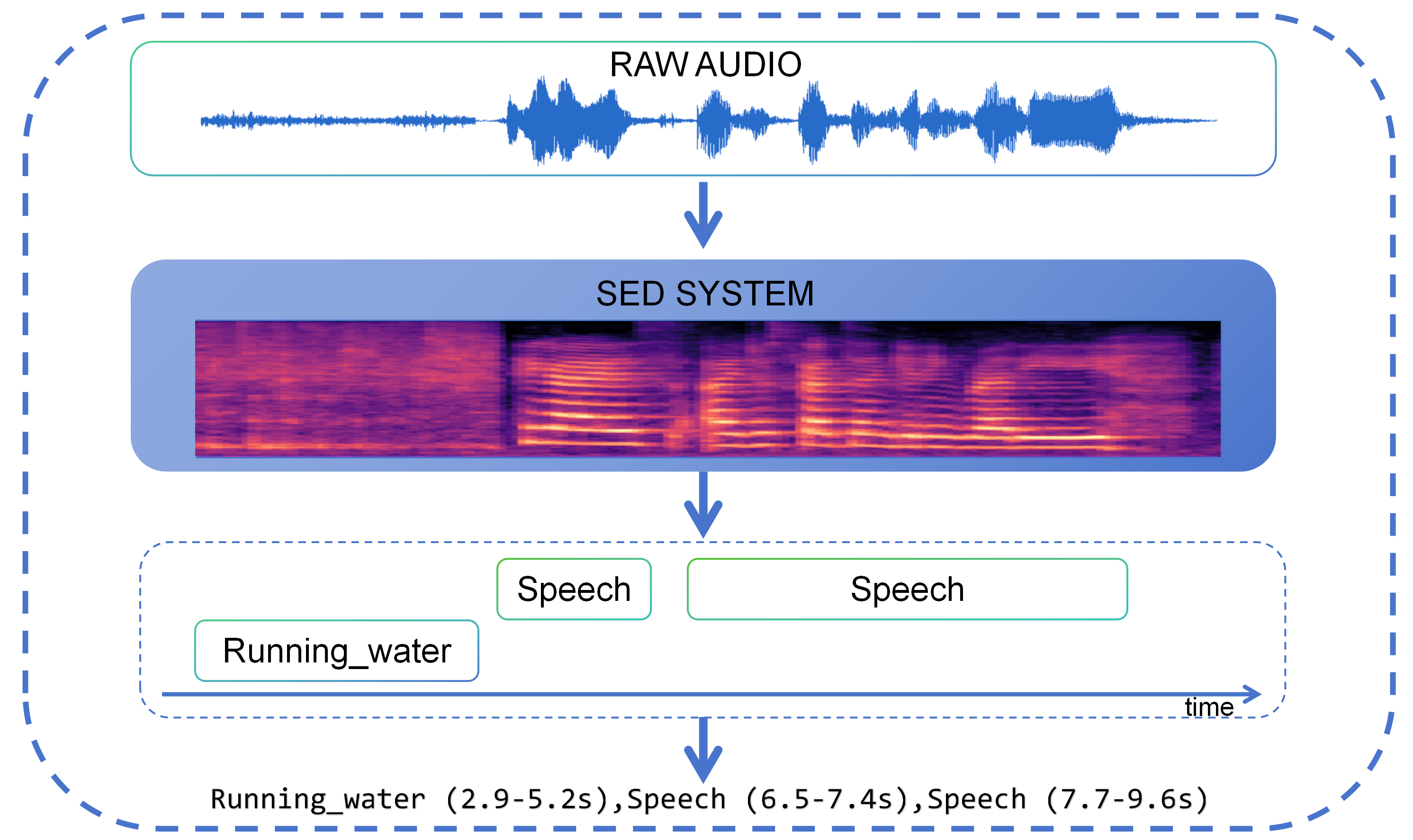}
    \caption{Illustration of a sound event detection system.}
    \label{SED}
\end{figure}

SED aims to identify the categories of acoustic events occurring within an audio recording and simultaneously determine their corresponding onset and offset timestamps within the temporal context. In contrast, the task of recognizing the presence of event categories without temporal localization is referred to as Audio Tagging (AT). Based on this distinction, SED can be viewed as a composition of two subtasks: a) \textit{Audio Tagging}, which focuses on detecting which event categories are present in the audio clip, and b) \textit{Temporal Boundary Detection}, which focuses on locating the precise time intervals during which these events occur. Fig.~\ref{SED} illustrates the operational differences between AT and SED. While AT produces a global classification vector (e.g., indicating ``speech" and ``running\_water" co-occurrence), SED requires fine-grained temporal resolution to detect event boundaries.

Formally, given an input audio sequence, $ X = \{x_1, x_2, \ldots, x_T\} $, consisting of $T$ temporal units (e.g., frames or time steps), the goal of AT is to predict a binary vector \( \mathbf\{p\} \in \{0,1\}^K \), where \( K \) denotes the number of predefined event classes, such that \( p_k=1 \) indicates the presence of class \( c_k \in \mathcal{C} \) in the entire segment. SED extends this formulation by requiring frame-level predictions that provide temporal localization for each event. Specifically, SED produces a sequence of predictions, \( \{y_t\}_{t=1}^T \) , where each \( y_t \in \mathcal{C} \cup {\emptyset} \) denotes either the presence of a particular event class \( c \in \mathcal{C} \) or the absence of any event (\( \emptyset \)) at time \( t \). Each detected event \( e \) is further characterized by its onset time \( t_{\text{on}}^{c} \) and offset time \( t_{\text{off}}^{c} \). Within the interval between these two timestamps, the model must consistently predict the corresponding event class \( c \). This constraint is formally expressed as: 
\begin{equation}
    y_t \in \left\{ c \;\middle|\; t \in \left[ t^{c}_{\text{on}},\; t^{c}_{\text{off}} \right] \right\} \\
    \label{eq:event_constraint}
\end{equation}

\subsection{CRNN-based Models}\label{subsec22}

In recent years, deep learning has emerged as the dominant paradigm in SED, with numerous architectures being explored and developed \citep{zhao2024sound}. Among them, convolutional recurrent neural networks (CRNNs) have shown strong performance, thanks to their ability to simultaneously capture local time-frequency features through convolutional layers and model long-term temporal dependencies via recurrent layers. Building upon this foundation, several notable enhancements have been proposed. For instance, the authors in \citep{singh2021polyphonic} introduced the mean-teacher CRNN (MT-CRNN), which employs a teacher-student framework to facilitate semi-supervised learning, significantly improving performance under limited supervision. Similarly, in the DCASE 2022 Challenge, the work in \citep{ebbers2022pre} achieved state-of-the-art results using forward-backward CRNNs (FBCRNNs) for weakly labeled and semi-supervised settings. For strongly labeled data, the authors in \citep{ebbers2022pre} adopted bidirectional CRNNs (Bi-CRNNs) to fully exploit bidirectional temporal context, effectively adapting their model to varying supervision scenarios. Beyond temporal modeling, advances have also been made in frequency-domain representation. The authors in \citep{nam2022frequency} proposed the frequency dynamic convolution (FDC) module, which introduces frequency-adaptive kernels to break the translational equivariance assumption of standard two-dimensional convolutions along the frequency axis. This design better aligns with the non-stationary characteristics of acoustic events, leading to improved recognition performance.

\subsection{SSL Models}
\label{subsec23}

Recent advances in self-supervised learning have significantly transformed audio representation learning by enabling models to extract rich acoustic patterns from large-scale unlabeled data. Wav2vec \citep{schneider2019wav2vec} pioneered unsupervised pre-training for speech recognition, demonstrating that leveraging unlabeled data can substantially reduce reliance on labeled resources. Wav2vec 2.0 \citep{baevski2020wav2vec} further improved this approach by introducing quantized latent representations as training targets, enabling more effective learning of continuous speech representations. HuBERT \citep{hsu2021hubert}, built upon the Transformer architecture, captures long-range dependencies via masked prediction, allowing the model to learn both structural and semantic aspects of speech. Based on HuBERT, WavLM \citep{chen2022wavlm} incorporates relative positional embeddings and simulates complex acoustic scenes through synthetic data, enhancing generalization to diverse real-world conditions. 

Beyond speech, self-supervised learning has also been explored in broader audio domains. PANNs \citep{kong2020panns} learn general-purpose audio features from large-scale data, while AST \citep{gong2021ast} applies the Transformer architecture to audio spectrogram patches, effectively capturing global spatiotemporal context. More recently, Dasheng \citep{dinkel2024dasheng} scales the MAE framework \citep{he2022masked} to 1.2 billion parameters, achieving strong results on audio-related tasks through large-scale pre-training. The authors in \citep{li2024self} proposed ATST-Frame, a frame-level teacher-student Transformer pre-training model fine-tuned on strongly-labeled AudioSet data \citep{shao2024fine}, achieving state-of-the-art performance.

\subsection{Post-processing}\label{subsec24}
Post-processing techniques in SED are designed to refine model outputs and enhance detection accuracy. These methods typically include filtering, threshold optimization, contextual integration, and class-specific adjustments, all of which contribute to improving the overall performance of SED systems. Among them, median filtering (MF) has been widely adopted to smooth frame-level predictions. However, fixed-window MF struggles to adapt to the diverse temporal characteristics of different sound events. To overcome this limitation, Liang et al. proposed the event-specific post-processing approach \citep{liang2022joint}, which analyzes high-confidence events in the development set to estimate the average duration for each category. This allows for adaptive window sizing per event class, significantly boosting detection performance. To further address the issue of temporal boundary degradation in frame-level thresholding, Ebbers et al. introduced the sound event bounding boxes (SEBBs) algorithm \citep{ebbers2024sound}. By decoupling duration estimation from confidence prediction, SEBBs reduce the interdependence between localization and classification reliability. Experimental results show that SEBBs significantly improved the polyphonic sound event detection scores (PSDS)  metric, setting a new benchmark for event-based evaluation in acoustic scene analysis.

Despite recent advancements, current SED systems still encounter two major limitations: 
\begin{itemize} 
\item \textbf{Insufficient Exploitation of SSL Diversity}: Most existing frameworks rely on a single self-supervised learning encoder, thereby overlooking the complementary representations that can be obtained from multiple heterogeneous SSL models. 
\item \textbf{Inflexible Post-processing}: Traditional post-processing techniques, such as median filtering, often rely on fixed parameters and therefore struggle to adapt to the varying temporal and structural characteristics of different sound events. As a result, they may lead to suboptimal performance in sound event detection.
\end{itemize}

To overcome these limitations, our approach introduces a multi-encoder fusion framework that integrates diverse SSL representations, along with an enhanced post-processing strategy based on normalized sound event bounding boxes (nSEBBs), thereby improving both event classification and temporal localization performance.

\section{Methods}
\label{sec3}

In this section, we provide a detailed explanation of the baseline and our proposed methods, which are designed to enhance sound event detection using self-supervised learning integration and adaptive post-processing. The CRNN+BEATs baseline architecture is presented in Section \ref{subsec31}. The SSL model fusion strategy is detailed in Section \ref{subsec32}, and Section \ref{subsec33} describes the normalized SEBBs post-processing framework.

\subsection{CRNN+BEATs Architecture}
\label{subsec31}

The CRNN+BEATs architecture \citep{dcase2023} has been taken as the baseline system in the DCASE 
Challenge 2023 and 2024 SED tasks, because of its state-of-the-art performance. 
This architecture leverages the pre-trained BEATs model, which currently achieves state-of-the-art 
results on the AudioSet classification task. Therefore, in this study, we also take 
CRNN+BEATs as our strong baseline system.

\begin{figure}
    \centering
    \includegraphics[width=0.8\linewidth]{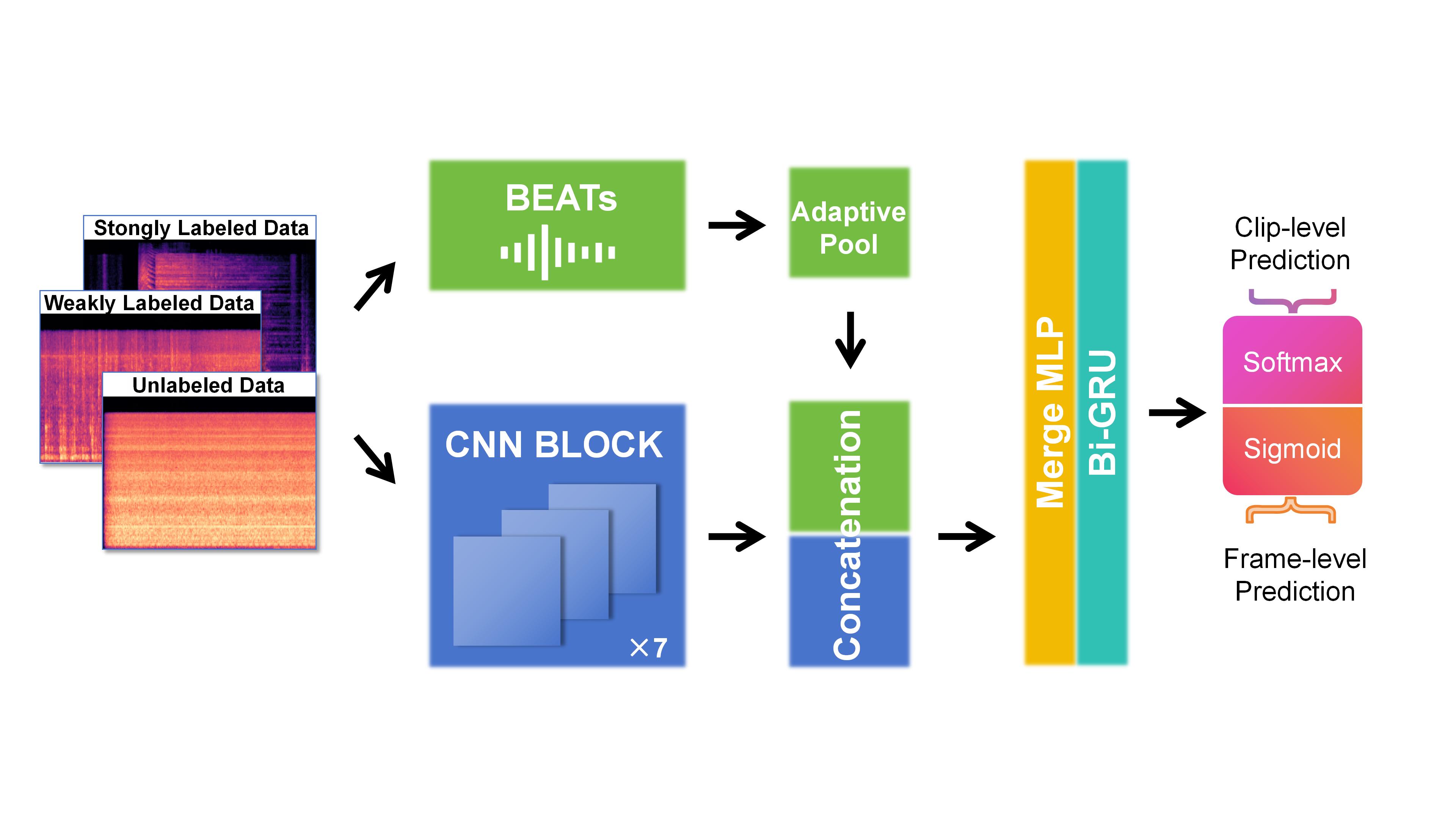}
    \caption{Architecture of the CRNN+BEATs baseline.}
    \label{DCASE}
\end{figure}

In the CRNN+BEATs model, the frame-level embeddings from pre-trained BEATs model are used in a 
late-fusion fashion with the existing CRNN classifier. Specifically, as shown in Fig.~\ref{DCASE}, during model training, each training batch includes audio clips of three types: strongly labeled, weakly labeled, and unlabeled. These clips are first transformed into Mel-spectrograms and passed through the BEATs model for feature extraction. The extracted features are then temporally aligned via adaptive average pooling and concatenated with CNN outputs. The combined representations are fed into a Merge MLP, followed by classification via the SED classifier. The classifier is capable of both frame-level and clip-level prediction. For frame-level prediction, it employs a sigmoid-based nonlinear normalization on the input features to generate frame-wise probabilities. For clip-level prediction, the frame-level probabilities are subsequently aggregated using temporal softmax attention weights. This attention mechanism helps preserve event-specific characteristics at each time frame while suppressing the influence of anomalous or irrelevant frames, thereby enhancing the robustness of the final clip-level output. 

\begin{figure}
    \centering
    \includegraphics[width=0.8\linewidth]{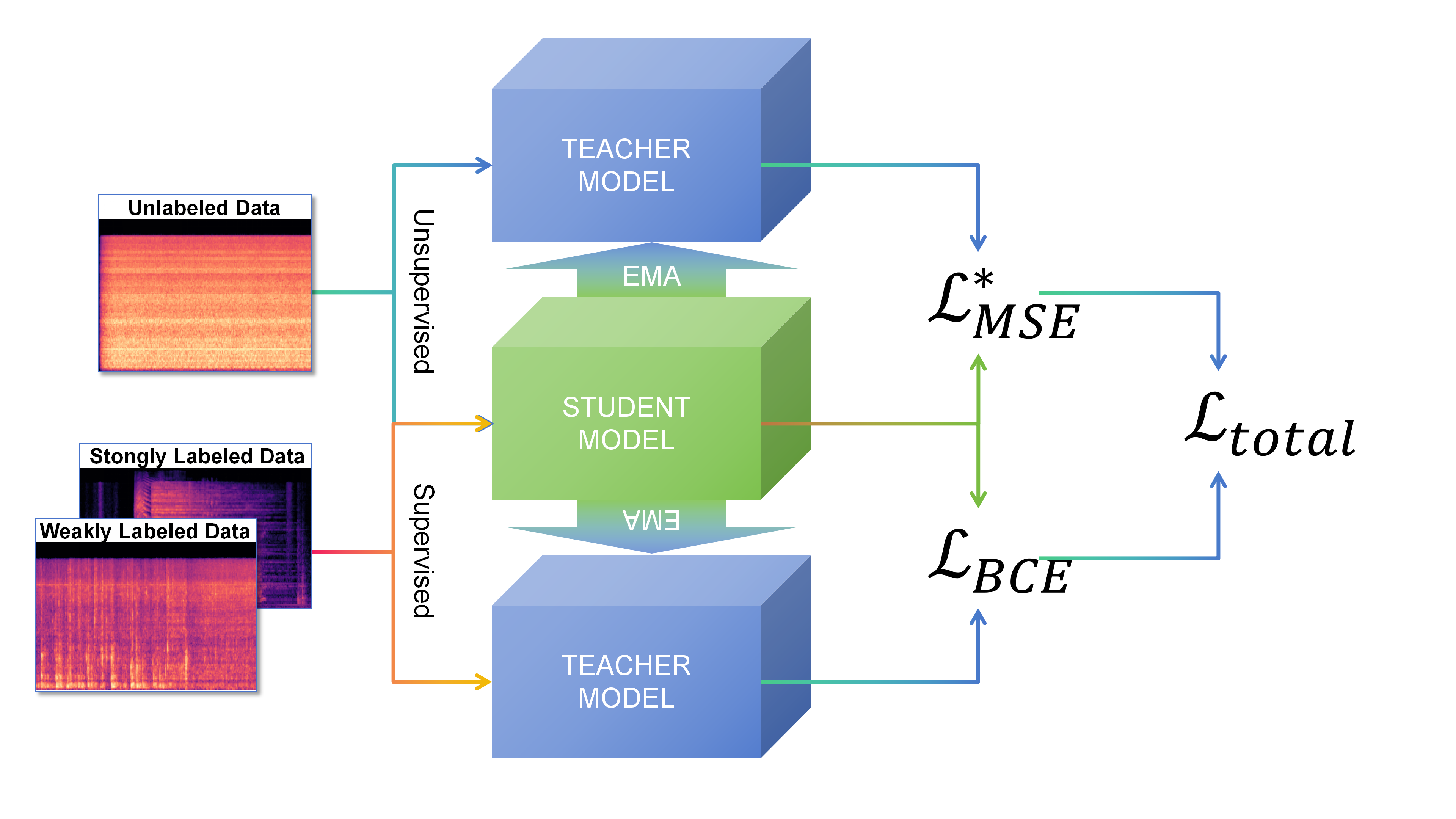}
    \caption{Loss calculation strategy for training the network (where the teacher model has the same structure, only the input data and the learning method used are different).}
    \label{LOSS}
\end{figure}

To enhance training stability and model robustness, the proposed system adopts a mean teacher\citep{tarvainen2017mean} learning framework. As clearly illustrated in Fig.~\ref{LOSS}, Binary Cross Entropy (BCE) and Mean Squared Error (MSE) are employed as the loss functions for supervised and self-supervised learning, respectively. During training, the teacher model is updated using the Exponential Moving Average (EMA) of the student model's parameters. During inference, both the student and teacher models are used for evaluation, and their performances are compared.

The overall training loss of the model is defined as follows:  
\begin{equation}
\mathcal{L}_{total} = \mathcal{L}_{BCE}^{weak} + \mathcal{L}_{BCE}^{strong} + \eta\cdot\left(\mathcal{L}_{MSE}^{weak} + \mathcal{L}_{MSE}^{strong} + \mathcal{L}_{MSE}^{unlabel} \right)
\end{equation}

For the labeled data, the discrepancy between predictions and ground truth is quantified using the binary cross-entropy (BCE) loss. For all data, including unlabeled samples, both the teacher and student models generate predictions, and a consistency loss is computed as the mean squared error (MSE) between these predictions. Importantly, the supervised loss is calculated solely based on the student model's outputs; the teacher model does not contribute to this step and is instead updated exclusively via exponential moving average (EMA) of the student model parameters~\citep{lawrance1977exponential}. Both the supervised loss $\mathcal{L}_{\text{BCE}}$ and the consistency loss $\mathcal{L}_{\text{MSE}}$ are computed at both the clip-level and the frame-level.

Here, $\mathcal{L}_{\text{BCE}}$ refers to the supervised binary cross-entropy loss, while $\mathcal{L}_{\text{MSE}}$ denotes the mean squared error that enforces consistency between the predictions of the teacher and student models. Loss terms with the superscript `weak' are computed using weakly labeled training data, those with the superscript `strong' are computed using strongly labeled data, and the loss with the superscript `unlabel' is derived from unlabeled data.

To balance the supervised and consistency losses, a weighting coefficient $\eta$ is introduced. This coefficient is gradually increased from 0 to a predefined maximum value (set to 2 in our implementation) following a warm-up schedule. This strategy allows the model to initially concentrate on learning from labeled data before progressively integrating information from unlabeled data.

\subsection{Integration and Fusion of SSL Representations}
\label{subsec32}

\begin{figure}
    \centering
    \includegraphics[width=1\linewidth]{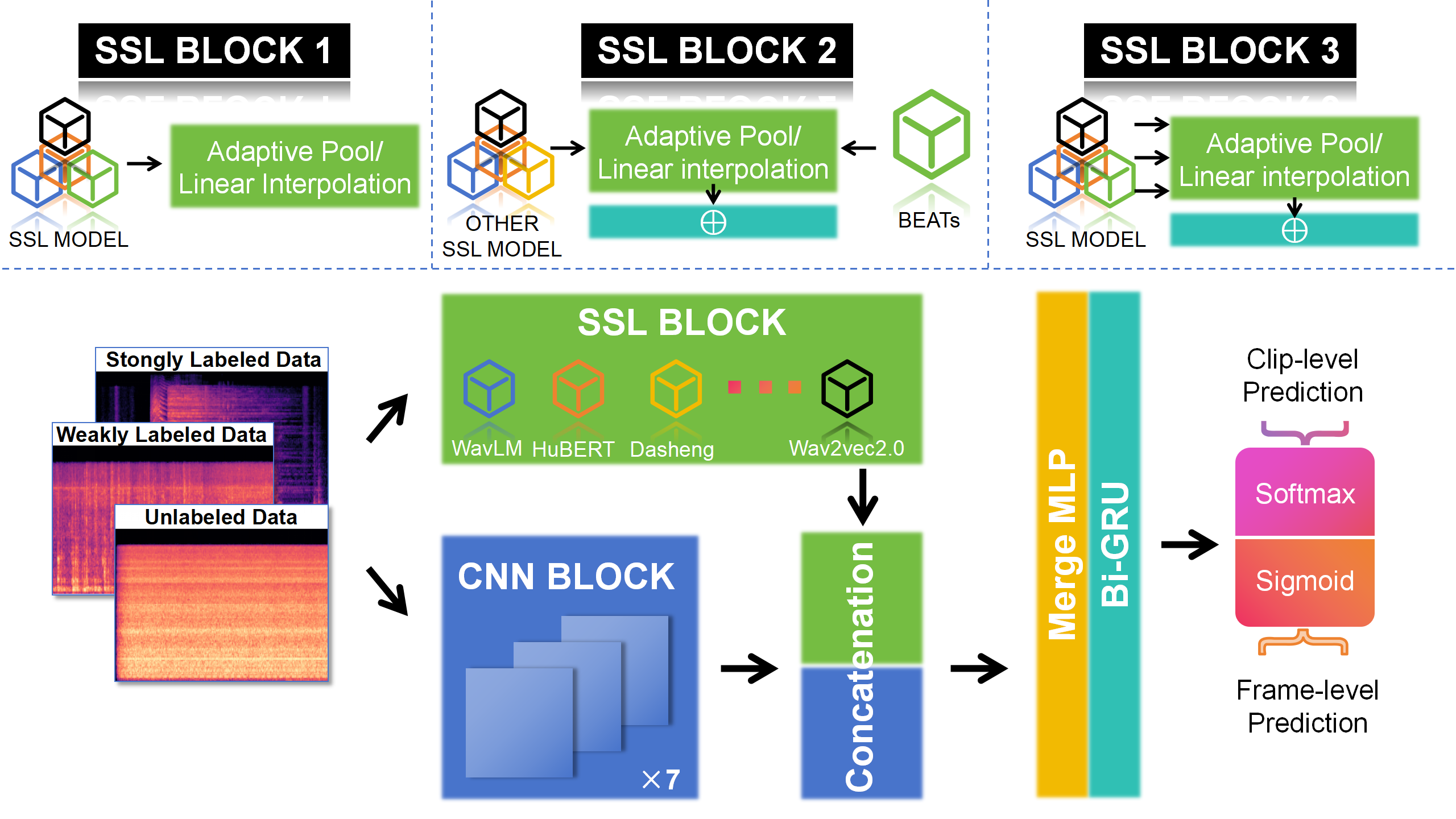}
    \caption{BLOCK 1: Individual SSL model. BLOCK 2: Dual-model fusion architecture. BLOCK 3: Full fusion model, where $\oplus$ denotes element-wise addition.}
    \label{FUSION}
\end{figure}

In this section, we explore strategies for combining diverse SSL representations to enhance 
complementary information integration and improve system robustness. Three 
SSL integration architectures are proposed as shown in Fig.~\ref{FUSION}: 

\begin{enumerate} [label=\arabic*)]
    \item \textbf{Individual SSL Representation Integration (SSL BLOCK 1):} We first explore the potential
    of using a different individual SSL model in SED,  by extracting audio representations independently from each pre-trained model (BEATs, HuBERT, wav2vec2.0, Dasheng, and WavLM). These representations are then taken as acoustic features and directly fed into the SED system to assess the representational quality and task-specific effectiveness of each SSL model.
    
    \item \textbf{Dual-Model Fusion (SSL BLOCK 2):} This architecture performs pairwise fusion of acoustic representations from BEATs and each of the other SSL models. Given that BEATs represents the current state-of-the-art in SED, we aim to enhance its performance by incorporating complementary information from other SSL models. To resolve mismatches in temporal resolution between feature extractors, we apply adaptive average pooling or linear interpolation to align feature dimensions before performing element-wise addition. This alignment ensures both effective integration and temporal consistency.
    
    \item \textbf{Full Fusion (SSL BLOCK 3):} To fully exploit the complementary strengths of all SSL models, we aggregate the features from all available models using element-wise addition. This fusion approach aims to maximize the synergistic potential of diverse audio representations and to explore the upper limits of performance achievable through multi-model integration.
\end{enumerate}

In addition to element-wise addition, we also explore feature concatenation as an alternative fusion strategy. Specifically, for SSL Block 2, we replace the addition operation with concatenation and subsequently fuse the combined features with CNN-derived features. Both fusion strategies (element-wise addition and concatenation) are experimentally compared to evaluate their respective effectiveness. 

\subsection{Normalized SEBBs Post-processing }
\label{subsec33}

In the work by \citep{ebbers2024sound}, the authors proposed Sound Event Bounding Boxes (SEBBs) as a post-processing strategy to enhance SED. Each predicted event is represented as a bounding box tuple:
\[
E_i = \{ c,t_{\text{on}}^c[i],\ t_{\text{off}}^c[i],\ S_c[i] \}
\]
where \( t_{\text{on}}^c[i] \) and \( t_{\text{off}}^c[i] \) denote the onset and offset timestamps of the \( i \)-th event of class \( c \), and \( S_c[i] \) represents a unified confidence score. This formulation explicitly decouples temporal boundary localization from confidence estimation, overcoming the limitations of traditional frame-level thresholding methods. By treating events as structured temporal objects rather than thresholded framewise outputs, SEBBs enable more accurate and interpretable SED results.
 To fully explore the potential of SEBBs, three variants were introduced 
in \citep{ebbers2024sound}: Threshold-based SEBBs (tSEBBs), Change-detection-based SEBBs (cSEBBs) and Hybrid SEBBs. 
Among these, experimental results showed that cSEBBs achieved the best performance.

The cSEBBs framework enhances event detection using confidence dynamics (delta scores $\Delta_c$) to identify event boundaries and merges adjacent segments via gap confidence comparisons, eliminating threshold-induced artifacts like boundary distortions and spurious splits. By aggregating frame-level confidence scores into event-level averages, it generates stable monotonic ROC curves for reliable threshold-based evaluation. The system employs three principal operational parameters: 1) the \textbf{step filter length ($l_{step} = \{0.32s, 0.48s, 0.64s\}$) }governs the temporal resolution of boundary analysis. It should be noted that the window size significantly affects the detection performance for different signal types; 
2) the step-wise dual merge thresholds-\textbf{ absolute ($\theta_{abs} = \{0.15, 0.2, 0.3\}$) and relative ($\theta_{rel}=\{1.5, 2.0, 3.0\}$) }- establish validity criteria for detected boundaries, where higher threshold values correspond to more stringent boundary confirmation requirements.

However, the cSEBBs framework relies on fixed parameter configurations determined through grid search, which introduces three key limitations: 1) lack of adaptability: static parameter ranges prevent context-sensitive tuning based on specific audio characteristics. 2) one-size-fits-all strategy: uniform parameters across all event classes ignore acoustic diversity. 3) high computational cost: exhaustive parameter search incurs exponential complexity, particularly in edge cases or under rare acoustic conditions.
\begin{algorithm}
\caption{Normalized cSEBBs Algorithm (nSEBBs)}
\label{alg:nsebbs}
\begin{algorithmic}[1]
\item \textbf{Input}: the raw frame-level score matrix of one audio recording with $T$ frames and $C$ target event classes: $S \in \mathbb{R}^{T \times C}$
\item \textbf{Output}: detected event boundaries $E$

\Procedure{nSEBBs}{$S$}
    \State $E \gets \emptyset$ \Comment{initialize event boundary set}
    \For{each sound class $c \in [1,C]$,}
        \State $s_c[t] \gets S[t,c], \ \ \forall t \in [1,T]$ \Comment{extract scores for class $c$ of frame $t$} \\ 
        \State //Calculate mean and variance
        \State $\mu_c = \mathrm{mean}\left( \{ s_c[t] \}_{t=1}^T \right), \ \sigma_c = \sqrt{ \mathrm{var}\left( \{ s_c[t] \}_{t=1}^T \right)}$ \\

        \State // Calculate PCR , avgED of each class  
        \State $\{{pcr}_{c},  {dur}_c \} \gets \text{CalStats}$($s_c[t], \mu_c, \sigma_c$)  
        \State $\{l_{step}, \theta_{rel}\}\gets \text{AdaptiveParameter}(pcr_c,dur_c)$ 
        \State $\hat{s}_c[t] \gets \text{StepFilter}(s_c[t], l_{step})$\Comment{apply step filtering}
        \State $\Delta_c[t] \gets |\hat{s}_c[t+1] - \hat{s}_c[t]|, \ \ \forall t \in [1,T-1]$ \Comment{calculate delta scores} \\
        
        \State // Detect boundaries from delta scores and merge close boundaries
        \State $\{B_c,S_{seg}^c\} \gets \text{FindBoundary}(\Delta_c[t],s_c[t])$ \Comment{find  change points}
        \State $E_i \gets \text{MergeSegments}(B_c, S^c_{\mathrm{seg}}, \theta_{\mathrm{rel}})$
        \State $E \gets E \cup E_i$
    \EndFor
    \State \Return $E$
\EndProcedure
\end{algorithmic}
\end{algorithm}

\begin{algorithm}
\caption{Calstats}
\label{alg:cal}
\begin{algorithmic}
\Procedure{Calstats}{$s_c[t], \mu_c, \sigma_c$}
    \State $nl \gets \text{Percentile}(s_c[t], 10)$
    \For{ each $t \in T$,}
    \If{ $s_c[t] > (\mu_c + 0.5\sigma_c)$ }
    \State $mask[t] = 1 $ \Else \ \ \ $mask[t] = 0 $ 
    \EndIf
    \EndFor \\
    \State $sl \leftarrow \frac{1}{T} \sum_{t=1}^T s_c[t] \cdot \mathrm{mask}[t] $ 
    \State $T_e \gets \sum_t mask[t]$,  $E_c \gets 0$, $in\_event \gets \textbf{false}$
    \For{each $t \in T$}
        \If{$mask[t] = 1$ and $in\_event = \textbf{false}$}
            \State $E_c \gets E_c + 1$
            \State $in\_event \gets \textbf{true}$
        \ElsIf{$mask[t] = 0$}
            \State $in\_event \gets \textbf{false}$
        \EndIf
    \EndFor
    \State \Return $\{10\log_{10}(sl / \max(nl, \epsilon)) , T_e / \max(E_c, \epsilon) \}$ \Comment{$\epsilon$ prevents division by zero} 
\EndProcedure
\end{algorithmic}
\end{algorithm}

\begin{algorithm}[H]
\caption{StepFilter: Compute Change Point Score}
\label{alg:step}
\begin{algorithmic}[1]
  \Procedure{StepFilter}{$s_c[t]$, $l_{step}$}
  \State $h =  l_{step}/2$
    \For{$t = h$ \textbf{to} $len(s_c) - h - 1$}
        \State $\hat{s}_c[t] = \frac{1}{h} \left(\sum_{i=t}^{t+h-1} s_c[i] - \sum_{i=t-h}^{t-1} s_c[i]\right)$
    \EndFor
    \State \Return $\hat{s}_c[t]$
  \EndProcedure
\end{algorithmic}
\end{algorithm}

\begin{algorithm}[H]
\caption{FindBoundary: Detect Event Boundaries}
\label{alg:find}
\begin{algorithmic}[1]
  \Procedure{FindBoundary}{$\Delta_c[t]$, $s_c[t]$}
    \State $t_{\text{on}}^c \gets [\ ]$, $t_{\text{off}}^c \gets [\ ]$ \Comment{event onset/offset candidates}
    \For{$t = 2$ \textbf{to} $len(\Delta_c) - 1$}
      \If{$\Delta_c[t] \leq \Delta_c[t{-}1]$ \textbf{and} $\Delta_c[t{-}1] > \Delta_c[t{-}2]$}
        \State $t_{\text{on}}^c$.append($t{-}1$) \Comment{local maximum: potential onset}
      \ElsIf{$\Delta_c[t] > \Delta_c[t{-}1]$ \textbf{and} $\Delta_c[t{-}1] \leq \Delta_c[t{-}2]$}
        \State $t_{\text{off}}^c$.append($t{-}1$) \Comment{local minimum: potential offset}
      \EndIf
    \EndFor
    \If{$len(t_{\text{on}}^c) > len(t_{\text{off}}^c)$}
      \State $t_{\text{off}}^c$.append($len(\Delta_c) - 1$) \Comment{add final offset if needed}
    \EndIf
    \State $B_c \gets [0]$ \Comment{start from audio beginning}
    \For{$i = 0$ \textbf{to} $len(t_{\text{on}}^c) - 1$}
      \State $B_c$.append($t_{\text{on}}^c[i]$)
      \State $B_c$.append($t_{\text{off}}^c[i]$)
    \EndFor
    \State $B_c$.append($len(\Delta_c) - 1$) \Comment{ensure last boundary is included}
    \State $S^c_{\mathrm{seg}} \gets [\ ]$
    \For{$j = 0$ \textbf{to} $len(B_c) - 2$}
      \State $start \gets B_c[j]$, $end \gets B_c[j+1]$
      \State $seg \gets s_c[start:end]$
      \State $S^c_{\mathrm{seg}}.\mathrm{append}((\mathrm{mean}(seg), \min(seg), \max(seg)))$ 
    \EndFor
    \State \Return $B_c$, $S^c_{\mathrm{seg}}$
  \EndProcedure
\end{algorithmic}
\end{algorithm}

\begin{algorithm}[H]
\caption{MergeSegments: Refine and Merge Segments}
\label{alg:merge}
\begin{algorithmic}[1]
 \Procedure{MergeSegments}{$B_c, S^c_{\mathrm{seg}}, \theta_{rel}$}
  \State {// At least two events in $B_c$ are required to trigger a merge segment}
  \If{$|B_c| \geq 6$} 
    \State $S_{max}^c[j] \gets \left\{ S^c_{\mathrm{seg}}[j][2] \mid j = 1,3,5,... \right\}$ \Comment{access max values}
    \State $S_{min}^c[j] \gets \left\{ S^c_{\mathrm{seg}}[j][1] \mid j = 2,4,6,... \right\}$ \Comment{access min values}
    \State $idx \gets \left\{ j \mid \frac{S_{max}^c[j]}{S_{min}^c[j]} < \theta_{rel} \land \frac{S_{max}^c[j+1]}{S_{min}^c[j]} < \theta_{rel} \right\}$
    \For{each $j$ in $idx$}
      \State Merge $B_c[2j+1]$ to $B_c[2j+2]$ into one segment
      \State Update $B_c$ and $S^c_{\mathrm{seg}}$ accordingly
    \EndFor
  \EndIf
  \State $t_{\text{on}}^c \gets B_c[1 :: 2]$ \Comment{odd-indexed boundaries}
  \State $t_{\text{off}}^c \gets B_c[2 :: 2]$ \Comment{even-indexed boundaries}
  \State $S_c \gets \left\{ S^c_{\mathrm{seg}}[j][0] \mid j = 1,3,5,... \right\}$ \Comment{use mean values}
  \State \Return $E_i = \{ c,t_{\text{on}}^c[i], t_{\text{off}}^c[i], S_c[i] \}$
 \EndProcedure
\end{algorithmic}
\end{algorithm}

To overcome these limitations, we propose a normalized sound event bounding boxes (nSEBBs) to enhance the cSEBBs, the whole procedure of nSEBBs is shown in Algorithm~\ref{alg:nsebbs}. It achieves content-adaptive parameter optimization through three statistical descriptors. While preserving the core computational architecture of cSEBBs – including delta score calculation ($\Delta_c$)(Algorithm~\ref{alg:step}), boundary localization(Algorithm~\ref{alg:find}), and segment merging(Algorithm~\ref{alg:merge}) – our method introduces two key innovations. We eliminate the static absolute merging threshold $\theta_{abs}$ while retaining the relative merging threshold $\theta_{rel}$, replacing fixed parameters with a signal-adaptive optimization process formalized in Algorithm~\ref{alg:cal}. This adaptation mechanism employs: 1) \text{Posterior contrast ratio (PCR)}, a novel confidence descriptor computed as the logarithmic ratio between  average of high posterior probabilities and the 10th percentile estimate; 2) \text{Average event duration (avgED)} derived from temporal mask analysis to control step filter length ($l_{step}$) allocation for transient/sustained sounds.

In Algorithm~\ref{alg:nsebbs}, the AdaptiveParameter subroutine is described as follows: the system ensures computational stability by pre-normalizing detection scores through statistical moment analysis, thereby maintaining numerical consistency within a controlled range. Unlike the fixed thresholding strategy employed in cSEBBs, our adaptive parameter tuning mechanism dynamically optimizes $\theta_{\text{rel}}$ via Equation~(\ref{eq:1}), in which the base value $\theta_{\text{base}}$ is modulated by weights derived from the PCR (Percentage of Coverage Rate). As shown in Table~\ref{tab:map}, this dual-faceted adaptation enables context-aware processing: under high PCR conditions, the merging threshold is tightened to preserve transient events, while prolonged avgED (average event duration) leads to an extended temporal filtering window to better capture sustained phenomena. The subsequent boundary detection and merging stages retain the structural consistency of cSEBBs but operate under these optimized parameters, effectively balancing detection sensitivity and over-segmentation across diverse acoustic environments.

\begin{equation}
\theta_{rel} = 
\begin{cases}
 \dfrac{pcr_c}{10} \times 0.1+\theta_{base}, & 0 \leq pcr_c < 10 \\
 \dfrac{pcr_c - 10}{20 - 10} \times 0.3+\theta_{base}, & 10 \leq pcr_c < 20 \\
 \dfrac{pcr_c- 20}{30 - 20} \times 0.3+\theta_{base}, & 20 \leq pcr_c< 30 \\
 \dfrac{pcr_c - 30}{50 - 30} \times 0.3+\theta_{base}, & 30 \leq pcr_c < 50
\end{cases}
\label{eq:1}
\end{equation}

\begin{table}[ht]
\centering
\renewcommand{\arraystretch}{1.5}
\caption{Corresponding mapping relationships.}
\label{tab:map}
\begin{tabular}{
  >{\centering\arraybackslash}m{3cm} 
  >{\centering\arraybackslash}m{2.5cm} 
  >{\centering\arraybackslash}m{3cm} 
  >{\centering\arraybackslash}m{2.5cm}
}
\toprule
\multicolumn{2}{c}{\textbf{Duration Parameters}} & \multicolumn{2}{c}{\textbf{Posterior Contrast Parameters}} \\
\cmidrule(r){1-2} \cmidrule(l){3-4}
$dur_c$ & $l_{step}$ & $pcr_c$ & $\theta_{base}$ \\
\midrule
$[0, 20)$   & 0.384 & $[0, 10)$   & 2.0 \\
$[20, 40)$  & 0.512 & $[10, 20)$  & 2.4 \\
$[40, 90)$  & 0.640 & $[20, 30)$  & 2.8 \\
$[90, 200)$ & 0.800 & $[30, 50)$  & 3.2 \\
\bottomrule
\end{tabular}
\end{table}
\begin{table}[htbp]
\centering
\renewcommand{\arraystretch}{2.0}
\caption{$\theta_{abs}$ mapping relationship in nSEBBs$_D$.}
\label{tab:mapd}
\begin{tabular}{
  >{\centering\arraybackslash}m{3cm}
  >{\centering\arraybackslash}m{2cm}
  >{\centering\arraybackslash}m{2cm}
  >{\centering\arraybackslash}m{2cm}
  >{\centering\arraybackslash}m{2cm}
}
\toprule
$dur_c$ & [0,20) & [20,40) & [40,90) & [90,200) \\
\midrule
$\theta_{\text{abs}}$ & 0.12 & 0.18 & 0.24 & 0.3 \\
\bottomrule
\end{tabular}
\end{table}

To further enhance the flexibility and robustness of our approach, we introduce a dual-threshold variant called nSEBBs$_D$. This variant simultaneously incorporates both relative ($\theta_{rel}$) and absolute ($\theta_{abs}$) thresholds, where the mapping relationship for $\theta_{abs}$ is detailed in Table~\ref{tab:mapd}. Except for this dual-threshold implementation, nSEBBs$_D$ maintains identical characteristics to the original nSEBBs framework. This configuration serves to systematically evaluate how different metrics perform when constrained by a mandatory dual-threshold mechanism. Similarly, as discussed in Section~\ref {sub56}, the cSEBBs$_D$ variant also results from the concurrent application of both thresholds. 
\begin{figure}
    \centering
    \includegraphics[width=1\linewidth]{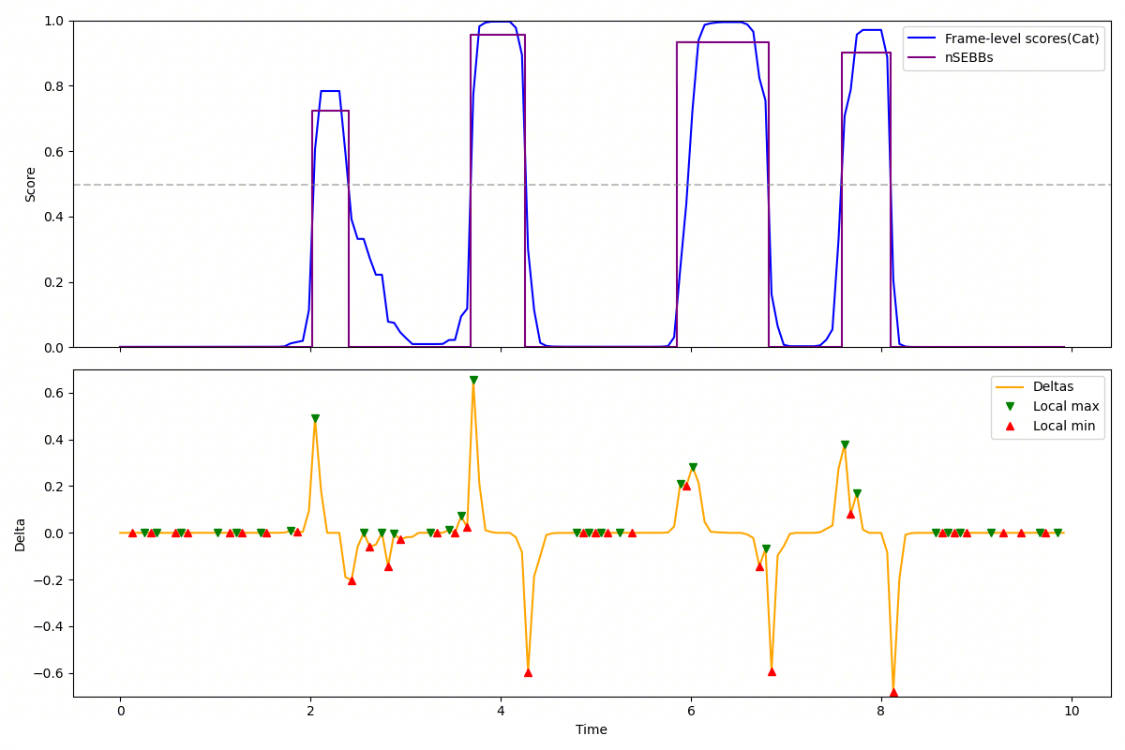}
    \caption{Proposed nSEBBs architecture's classification output for the 'Cat' category audio sample (Y\_\_p-iA312kg\_70.000\_80.000.wav). }
    \label{nSEBB}
\end{figure}  
Moreover,Fig.~\ref{nSEBB} illustrates the boundary selection strategy of the proposed nSEBBs framework. The frame-level prediction scores, generated by a CRNN+BEATs model for the cat-class audio segment *Y\_\_p-iA312kg\_70.000\_80.000.wav*, are processed to compute delta scores (Algorithm~\ref{alg:nsebbs}). This enables precise identification of boundary onset points, followed by segment merging to produce the final nSEBBs.

\section{Experimental Configurations}\label{sec4}

\subsection{Dataset}\label{subsec41}
The dataset used in this study is derived from the development set of Task 4 in the DCASE 2023 Challenge \citep{turpault2019sound}. This dataset, known as \textbf{DESED} (Sound Event Detection in Domestic Environments), adopts a multi-modal annotation scheme and includes three main types of data: 1) \textbf{Strongly labeled data}, annotated with both event categories and precise temporal boundaries; 2) \textbf{Weakly labeled data}, annotated only with event categories, without temporal information; 3) \textbf{Unlabeled data}, containing no annotations.

The training set includes 1,578 weakly labeled audio clips, 14,412 unlabeled clips, 3,470 real-world strongly labeled samples, and 10,000 synthetically generated strongly labeled samples. This hybrid annotation strategy facilitates effective learning by leveraging both real and synthetic data while minimizing manual labeling effort. The validation set comprises 1,168 strongly labeled clips, each annotated with complete temporal information, and is curated under strict labeling protocols. It serves as a robust benchmark for assessing model performance. All experiments are evaluated on the validation set.

\subsection{Models}\label{subsec42}

Five pre-trained SSL models with different architectures are adopted in our study: 

\textbf{Wav2vec 2.0} \citep{baevski2020wav2vec}: It is a widely used SSL model for speech representation learning. It captures contextualized representations from raw, unlabeled audio through a combination of contrastive learning and quantization techniques. With only a small amount of labeled data, wav2vec 2.0 can be fine-tuned to achieve strong performance, showing great potential for SED tasks where annotated resources are limited.

\textbf{HuBERT} \citep{hsu2021hubert}: It builds upon the SSL paradigm by introducing an offline clustering process to generate discrete pseudo-labels, which serve as targets for masked prediction during pre-training. Through iterative refinement of these cluster assignments, HuBERT is able to learn high-quality speech representations and demonstrates strong generalization to a wide range of downstream tasks.

\textbf{WavLM} \citep{chen2022wavlm}: It is an advanced SSL model designed for full-stack speech processing. It enhances the masked speech modeling framework by incorporating a denoising objective and a gated relative position bias, which improves its ability to handle noisy inputs and model long-range dependencies. In this study, we adopt both the WavLM$_{base}$ and $\mathrm{WavLM}_{base+}$ variants, which differ in terms of the amount and diversity of their pretraining data.

\textbf{BEATs} \citep{chen2022beats}: It is a transformer-based SSL model that focuses on learning high-level audio semantics via discrete label prediction. By modeling masked audio segments with semantic supervision, BEATs achieves competitive results on benchmarks such as AudioSet-2M and ESC-50. We utilize the $\mathrm{BEATs}_{iter3+}$ (AS2M) variant in our system for extracting semantic audio features.

\textbf{Dasheng} \citep{dinkel2024dasheng}: It is a scalable masked autoencoder-based audio encoder designed for general-purpose audio classification. It employs an efficient masking and reconstruction strategy to learn robust latent representations and has shown excellent performance on the HEAR benchmark \citep{turian2022hear}. In our system, we use the $\mathrm{Dasheng}_{base}$ model for audio embedding extraction.

With all these SSL models, When processing 10-second audio segments, these SSL models generate embeddings with different temporal resolutions: Dasheng outputs feature sequences of length $T=250$, $D=768$; BEATs yields $T=496$, $D=768$; and WavLM, HuBERT, and wav2vec 2.0 each produce embeddings of $T=499$, $D=768$. 

\subsection{Configurations}\label{subsec43}

All audio signals were standardized through downsampling from 44 kHz to 16 kHz. To enhance data diversity, both weak and synthetic samples were augmented via batch-wise mixup (linear interpolation of input features). Each experiment was repeated three times with different random seeds to ensure reproducibility. 

The architecture of CNN BLOCK, MLP and BiGRU blocks are exactly the same as the 
CRNN+BEATs model that used as the baseline system in DCASE 2023 Challenge Task4 \citep{turpault2019sound}. 
All experiments utilized a single NVIDIA GeForce RTX 2080 Ti GPU and employed the Adam optimizer with consistent hyperparameters across models: a batch size configuration of [24,24,48], initial learning rate of 1e-3, 50-epoch warm-up period, 200 total training epochs, EMA decay rate of 0.999, and dropout probability of 0.5. For post-processing, a 7-point median filter was applied uniformly. To further enhance system performance, both cSEBBs and nSEBBs were applied in our experiments to see their effectiveness.

\subsection{Evaluation Metrics}\label{subsec44}

Three types of evaluation metics are used to measure our SED system performance, they are the PSDS \citep{bilen2020framework}, the intersection-based F$_1$ (inter-F$_1$) score and the event-based F$_1$ (event-F$_1$) score. Specifically, $\mathrm{PSDS}_1$ $(\rho_{DTC} = \rho_{GTC} = 0.7, \alpha_{CT} = 0, \alpha_{ST} = 1, \mathrm{eFPR_{max}} = 100/\mathrm{h})$ emphasizes precise event boundary localization. $\mathrm{PSDS}_2$ $(\rho_{DTC} = \rho_{GTC} = 0.1, \rho_{CTCC} = 0.3, \alpha_{CT} = 0.5, \alpha_{ST} = 1, \mathrm{eFPR_{max}} = 100/\mathrm{h})$ focuses on classification accuracy, penalizing confusion and ghost events. Complementarily, the inter-F$_1$ is computed with the following criteria: a detection tolerance criterion (DTC) and ground truth intersection criterion (GTC) threshold of 0.5, alongside a cross-trigger tolerance criterion (CTTC) threshold of 0.3. The inter-F$_1$ metric serves to quantitatively evaluate the spatial-temporal alignment between predicted and ground-truth sound event labels, offering a robust assessment of localization accuracy. The event-F$_1$ score provides a stricter metric by requiring both correct class prediction and temporal alignment within tolerance windows. Higher values indicate better preservation of event identity and timing, which is critical for time-sensitive applications such as surveillance or alarm systems.

\section{Results}\label{sec5}
\subsection{Baselines}\label{sub51}

Table~\ref{tab:base} presents the results of the baseline model (MT-CRNN) and its BEATs-enhanced variant (CRNN+BEATs), alongside several recent top methods proposed in the literature. Our re-implementation of MT-CRNN and CRNN+BEATs yielded minor discrepancies from official results, attributable to hardware variations (e.g., GPU specifications). These include: 1) FDY-CRNN \citep{nam2022frequency}, which eliminates spectral translation equivariance via frequency-dynamic convolutions with adaptive kernels; 2) ASiT-CRNN \citep{zheng2025asit}, which improves multi-scale feature learning through local-global spectrogram transformers integrated with CRNN; 3) FDY-LKA-CRNN \citep{kim2023semi}, which adopts a semi-supervised mean-teacher framework, combining frequency-aware convolutions and BEATs embeddings to address label scarcity, achieving SOTA performance in DCASE 2023 Task4a; 4) MAT-SED \citep{cai2024mat}, a transformer-based model trained with masked spectrogram reconstruction for temporal-semantic optimization; 5) ATST-SED \citep{shao2024fine}, which leverages ATST-Frame for fine-grained audio representations and achieves SOTA results.
\begin{table}[!ht]
    \centering
    \renewcommand{\arraystretch}{1.3}
    \caption{Performance comparison of CRNN-based  and other competitive models on DCASE 2023 Task4 validation set.}
    \label{tab:base}
    \begin{tabular}{cccc}
        \toprule
        \textbf{Model} & \textbf{PSDS}$_1$ & \textbf{PSDS}$_2$ & \textbf{event-F}$_{1}$(\%)\\
        \midrule
        MT-CRNN & 0.353 & 0.522 & 40.8 
\\
        CRNN+BEATs & 0.497 & 0.727 & 55.7 \\
        CRNN+BEATs(cSEBBs) & 0.511 & - & - \\
        \midrule
        FDY-CRNN \citep{nam2022frequency}& 0.452 & 0.670 & 53.3 \\
        ASiT-CRNN \citep{zheng2025asit}& 0.488 & 0.767 & 56.1 \\
        FDY-LKA-CRNN \citep{kim2023semi}& 0.525 & 0.775 & 62.9 \\
        MAT-SED \citep{cai2024mat}& 0.587 & 0.792 & - \\
        ATST-SED \citep{shao2024fine}& 0.583 & 0.810 & 63.4 \\
        \bottomrule
    \end{tabular}
\end{table}

Among all these results in Table~\ref{tab:base}, MAT-SED and ATST-SED show superior overall performance across the key metrics. Notably, ATST-SED achieves the highest event detection performance with an event-F$_1$ of 63.4 \% while maintaining competitive results in PSDS$_1$ (0.583) and PSDS$_2$ (0.810). On the other hand, MAT-SED achieves the top PSDS$_1$ of 0.587, further underscoring its strong detection capabilities. These two models currently represent the best-performing approaches in this domain. And we also see that 
FDY-LKA-CRNN with BEATs feature embedding integration also achieves good results. All these results tell us 
that pre-trained strong SSL models can provide significant performance gains in SED systems.

Moreover, compared to the MT-CRNN baseline, models that incorporate BEATs representations, CRNN+BEATs and CRNN+BEATs(cSEBBs), and even FDY-LKA-CRNN system, demonstrate substantial gains in both  PSDS$_1$ and event-F$_1$. This indicates that BEATs embeddings significantly enhance the feature representations of conventional CRNN-based system. In contrast, models without such enhancements (e.g., MT-CRNN, FDY-CRNN, and ASiT-CRNN) generally show lower performance, particularly in the event-based \( F_1 \)-score, trailing behind by more than 10 percentage points. Overall, these results highlight the effectiveness of incorporating well pre-trained SSL representations (such as BEATs and ATST) in pushing the frontier of SED performance. 
As our goal is to investigate the effectiveness of SSL representations rather than pursuing  SOTA performance, we adopt CRNN+BEATs as our baseline model. 
\subsection{Results with SOTA SSL Models}\label{sub52}
\subsubsection{Results with individual SSL representation integration}

\begin{table}[t]
    \centering
    \caption{Performance comparison of CRNN-based models with different individual SSL representations.}
    \label{tab:inssl}
    \renewcommand{\arraystretch}{1.3} 
    \begin{tabular}{lcccc}
        \toprule
        \textbf{Model} & \textbf{PSDS$_1$} & \textbf{PSDS$_2$} & \textbf{event-F$_1 (\%)$}& \textbf{inter-F$_1 (\%)$}\\
        \midrule
        MT-CRNN& 0.353 & 0.522 & 40.8 & 62.4 \\
        CRNN+BEATs& 0.497 & 0.727 & 55.7 & 79.0 \\
        CRNN+wav2vec2.0$_{\text{base}}$ & 0.350 & 0.516 & 42.9 & 64.1 \\
        CRNN+HuBERT$_{\text{base}}$ & 0.352 & 0.520 & 43.7 & 63.5 \\
        CRNN+Dasheng$_{\text{base}}$ & 0.419 & 0.626 & 50.0 & 70.9 \\
        CRNN+WavLM$_{\text{base}}$ & 0.342 & 0.506 & 39.4 & 61.5 \\
        CRNN+WavLM$_{\text{base+}}$ & 0.342 & 0.516 & 40.3 & 62.4 \\
        \bottomrule
    \end{tabular}
\end{table}

Table \ref{tab:inssl} shows the performance comparison of CRNN-based models with different 
individual SSL representations as shown in SSL BLOCK 1 in Fig.~\ref{FUSION}. 
The baseline MT-CRNN achieves PSDS${_1}$/PSDS${_2}$ scores of 0.353/0.522 and an inter-F$_1$ of 62.4\%.
Among all models, CRNN+BEATs achieves the best performance, significantly improving both PSDS (relative 40.8\% 
PSDS${_1}$ and 39.3\% PSDS${_2}$ gains) and inter-F$_1$ (relative 16.6\% improvement), demonstrating strong feature 
complementary information capabilities with CNN outputs in the CRNN backbone. In contrast, speech-oriented models show varied results. CRNN+Dasheng$_\text{base}$ yields moderate improvements (PSDS$_1$ = 0.419, inter-F$_1$ = 70.9), while wav2vec2.0, HuBERT, and WavLM offer only marginal gains, particularly in PSDS, suggesting 
different pre-trained SSL models show different acoustic representation ability, because of their different model architecture and training datasets, and thus showing different complementary information with CNN 
outputs in the whole CRNN+* architecture. These findings identify BEATs as the most promising SSL 
representation for feature shallow fusion, with Dasheng as a secondary candidate. This motivates the hierarchical fusion strategy employed in SSL BLOCKS 2–3 in  Fig.~\ref{FUSION}, where BEATs acts as the primary feature integrator.

\subsubsection{Results with dual-model and full fusion}

Table \ref{tab:SSL2} presents the performance of CRNN with different SSL representation 
integration methods. As shown in Fig.~\ref{FUSION}, two embedding fusion methods are 
explored: embedding concatenation and direct addition. From these results, we see that 
augmenting the CRNN+BEATs baseline with additional SSL embeddings generally had only modest effects. 
The baseline (CRNN+BEATs alone) achieves PSDS$_1$ = 0.497, PSDS$_2$ = 0.727, event-F$_1$ = 55.7\%, 
inter-F$_1$ = 79.0\%. Almost all SSL-augmented models produced lower PSDS$_1$ (no model exceeded 0.497), 
and only one configuration (WavLM$_{base}$ with additive fusion) slightly raised PSDS$_2$ (0.729 vs 0.727). 
In contrast, some models did improve detection F$_1$: for example, WavLM$_{base}$ (concat) yields the highest 
event-F$_1$ (58.6\%) and inter-F$_1$ (80.2\%), surpassing the baseline by around 2.9 and 1.2 points, 
respectively, whereas wav2vec 2.0$_{base}$ (add) gave a smaller event-F$_1$ boost (56.3\%). Other SSL additions 
(HuBERT$_{base}$, Dasheng$_{base}$) tended to stay near or below baseline F$_1$. The ``Full Fusion” model 
(combining all SSLs) performed worst on all metrics (PSDS$_1$ = 0.448, event-F$_1$ = 53.7\%), suggesting that 
naive fusion of all representations can harm performance. Regarding fusion strategy, additive fusion generally 
preserved higher PSDS scores (closer to baseline) and often gave slightly better F$_1$ for speech-pre-trained 
models, while concatenation sometimes raised F$_1$ at the expense of PSDS. For instance, wav2vec2.0 and HuBERT 
variants achieved higher PSDS$_1$ and event-F$_1$  with addition than concatenation, whereas for WavLM variants 
concatenation produced larger  F$_1$ gains (with add yielding a marginal PSDS$_2$ increase). Overall, the most 
significant improvements over baseline came from WavLM-based models: WavLM$_{base}$ (concat) gave the highest F$_1$ score, 
and WavLM$_{base}$ (add) achieved the best PSDS$_2$. In summary, integrating SSL representations provided only 
incremental benefits: the best F$_1$ increase was a few percentage points, and PSDS metrics mostly declined.

\begin{table}[ht]
\centering
\renewcommand{\arraystretch}{1.3}
\caption{Performance of CRNN with different SSL representation integration methods.}
\label{tab:SSL2}
\setlength{\tabcolsep}{6pt}
\begin{tabular}{@{}lllcccc@{}}
\toprule
\multirow{2}{*}{Function}& \multirow{2}{*}{Model} & \multirow{2}{*}{Fusion} & \multicolumn{4}{c}{Metrics} \\
\cmidrule(l){4-7} 
 & & &\textbf{PSDS$_1$} & \textbf{PSDS$_2$} & \textbf{event-F$_1 (\%)$}& \textbf{inter-F$_1 (\%)$}\\
\midrule
BLOCK 1& CRNN+BEATs & - & 0.497 & 0.727 & 55.7 & 79.0 \\
\midrule
\multirow{10}{*}{BLOCK 2}& \multirow{2}{*}{CRNN+BEATs+wav2vec 2.0$_{base}$}& add & 0.471 & 0.715 & 56.3 & 78.1 \\ 
& & concat & 0.463 & 0.699 & 54.9 & 77.0 \\ 
& \multirow{2}{*}{CRNN+BEATs+HuBERT$_{base}$}& add & 0.475 & 0.711 & 55.2 & 78.0 \\ 
& & concat & 0.464 & 0.689 & 54.9 & 77.1 \\ 
& \multirow{2}{*}{CRNN+BEATs+Dasheng$_{base}$}& add & 0.452 & 0.675 & 56.0 & 75.6 \\ 
& & concat & 0.451 & 0.684 & 55.3 & 76.8 \\ 
& \multirow{2}{*}{CRNN+BEATs+WavLM$_{base}$}& add & 0.484 & \textbf{0.729} & 56.5 & 78.4 \\ 
& & concat & 0.474 & 0.712 & \textbf{58.6} & \textbf{80.2} \\ 
& \multirow{2}{*}{CRNN+BEATs+WavLM$_{base+}$}& add & 0.477 & 0.721 & 55.8 & 78.1 \\ 
& & concat & 0.477 & 0.708 & 57.3& 79.7\\ 
\midrule
BLOCK 3&Full Fusion& add & 0.448 & 0.689 & 53.7 & 75.7 \\ 
\bottomrule
\end{tabular}
\end{table}

Notably, all these performance difference between using different SSL models 
to extract acoustic embeddings may due to the differ in their model structures and 
pretraining corpus. Dasheng was trained on large-scale general audio datasets (including AudioSet), 
whereas wav2vec 2.0 and HuBERT were trained on speech (LibriSpeech), and WavLM on a 
very large noisy English speech collection (around 94k hours). 
Despite Dasheng’s audio-domain pretraining, its embeddings did not yield clear advantages: Dasheng$_{base}$ (add) 
improved event-F$_1$ only modestly (56.0\%) and had among the lowest PSDS scores, implying limited complementary 
benefit. In contrast, the WavLM embeddings (speech-trained) delivered the largest F$_1$ gains, suggesting that 
its extensive speech pretraining captured useful general-purpose features. Thus, the results suggest that the 
CRNN+BEATs baseline already encodes much of the relevant information, and adding SSL features, whether from 
speech or audio domains, yields at best marginal gains. Fusion strategy (add vs. concat) had some impact on 
specific metrics but did not dramatically change overall trends: additive fusion tended to better preserve 
baseline detection scores, while concatenation sometimes boosted F$_1$ at the cost of PSDS. In conclusion, 
combining CRNN+BEATs with SSL representations (especially WavLM) can slightly improve event detection F$_1$, but 
no configuration substantially outperforms the strong baseline, and the expected benefits from different 
pretraining domains (speech vs. audio) appear to be limited.

\begin{sidewaystable}
\caption{Event-based F$_1$(event-F$_1$) scores for the ten categories on DCASE 2023 Task 4 validation set.}
\label{data4}
\centering 
\begin{tabular}{l *{10}{c}} 
\toprule
\textbf{Model} & \textbf{Alarm} & \textbf{Blender} & \textbf{Cat} & \textbf{Dishes} & \textbf{Dog} & \textbf{Electric} & \textbf{Frying} & \textbf{Water} & \textbf{Speech} & \textbf{Vacuum} \\ 
& \textbf{bell} & & & & & \textbf{shaver} & & \textbf{running} & & \textbf{cleaner} \\ 
\midrule
MT-CRNN & 46.4 & 41.0 & 44.2 & 27.2 & 25.5 & 45.2 & 33.1 & 36.3 & 53.2 & 56.7 \\ 
\midrule
CRNN+BEATs & \textbf{53.7} & 62.2 & \textbf{54.5} & 34.4 & 36.6 & 70.0 & 58.6 & 57.1 & \textbf{69.9} & 72.4 \\ 
CRNN+wav2vec2.0$_{\text{base}}$ & 47.9 & 34.5 & 43.3 & 21.4 & 30.5 & 56.9 & 35.8 & 42.0 & 57.7 & 59.9 \\ 
CRNN+HuBERT$_{\text{base}}$ & 50.0 & 41.9 & 45.6 & 20.3 & 25.3 & 59.3 & 32.7 & 38.7 & 59.6 & 64.5 \\ 
CRNN+Dasheng$_{\text{base}}$ & 49.5 & 50.2 & 43.5 & 24.6 & 30.1 & 69.1 & 41.6 & 49.7 & 67.4 & 74.6 \\ 
CRNN+WavLM$_{\text{base}}$ & 46.8 & 39.1 & 44.0 & 16.4 & 28.2 & 45.3 & 32.4 & 38.9 & 59.8 & 44.1 \\ 
CRNN+WavLM$_{\text{base+}}$ & 45.9 & 38.1 & 43.0 & 19.3 & 25.9 & 50.0 & 35.7 & 39.7 & 59.8 & 45.7 \\ 
\midrule
CRNN+BEATs+wav2vec2.0$_{\text{base}}$ & 52.2 & 57.1 & 50.8 & 33.2 & \textbf{37.4} & 69.6 & 61.4 & 57.3 & 66.5 & \textbf{77.7} \\ 
CRNN+BEATs+HuBERT$_{\text{base}}$ & 51.2 & 53.4 & 51.2 & 32.5 & 36.6 & 70.5 & 57.8 & \textbf{60.9} & 68.6 & 69.5 \\ 
CRNN+BEATs+Dasheng$_{\text{base}}$ & 52.8 & \textbf{65.6} & 48.2 & 27.9 & 30.9 & 75.0 & 57.6 & 58.8 & 67.8 & 75.9 \\ 
CRNN+BEATs+WavLM$_{\text{base}}$ & 53.6 & 60.8 & 54.4 & 34.4 & 35.7 & \textbf{76.1} & 66.0 & 57.3 & 69.6 & 57.4 \\ 
CRNN+BEATs+WavLM$_{\text{base+}}$ & 51.2 & 60.6 & 53.1 & \textbf{36.3} & 37.1 & 68.1 & 63.4 & 57.3 & 69.1 & 57.0 \\
\midrule
Full Fusion & 52.7 & 52.8 & 51.0 & 28.3 & 35.6 & 53.1 & \textbf{66.7} & 56.3 & 66.8 & 74.7 \\ 
\bottomrule
\end{tabular}
\end{sidewaystable}

\subsection{Class-wise Results with SSL models}

We also perform a class-wise analysis to evaluate how SSL models enhance category-specific performance. As detailed in Table \ref{data4}, speech-pretrained SSL models (wav2vec2.0, HuBERT, WavLM) exhibit strong specialization in speech-related categories but limited generalization to non-vocal sounds. While these models outperform the MT-CRNN baseline in the speech category (F$_1$: 57.7\%–59.8\% vs. 53.2\%), their performance degrades 
in non-speech events (e.g., WavLM$_{\text{base}}$ achieves only 16.4\% F$_1$ for Dishes vs. MT-CRNN’s 27.2\%). 
However, combining these speech-pretrained SSL representations with BEATs embedding eliminates this 
imbalance: For example, BEATs+WavLM$_\text{base}$ boosts Dog detection to 35.7\% F$_1$ (+10.2 over WavLM alone) while preserving speech performance (69.6\% F$_1$). This feature fusion between two different SSL representations extends across domains: BEATs+wav2vec2.0 improves Frying to 61.4\% F$_1$ (+25.6 over standalone wav2vec2.0), and 
BEATs+Dasheng achieves state-of-the-art performance on Blender (65.6\% F$_1$). However, not all fusions are 
beneficial across the board. Dasheng-based combinations underperform in categories like Cat (48.2\% vs. BEATs’ 54.5\%), suggesting that domain-specific SSL models may introduce conflicting inductive biases despite their shared learning objectives. The Full Fusion configuration shows extreme variance, excelling in Frying (66.7\% F$_1$) but collapsing in Electric\_shaver (53.1\% vs. BEATs’ 70.0\%), indicating over-aggregation risks. These results demonstrate that targeted SSL fusion—rather than universal integration—optimally balances speech-specific and general acoustic event detection.

\subsection{Results with SEBBs-based Post-processing}\label{sub56}

In this section, we provide a detailed comparative analysis of various SEBBs-based post-processing methods and their performance outcomes. The proposed nSEBBs strategy offers substantial computational efficiency while maintaining competitive detection accuracy, as shown in Tables~\ref{data5} and~\ref{data6}. On an Intel Core i7-11800H platform, nSEBBs reduces the average runtime by 93.8\% compared to the conventional cSEBBs method (8s vs. 130s). Despite this drastic speed-up, nSEBBs delivers comparable improvements in PSDS$_1$ for CRNN integrated with individual SSL models, for example, as shown in Table \ref{data5}, WavLM$_\text{base+}$ achieves 0.366 with nSEBBs versus 0.379 with cSEBBs, and HuBERT$_\text{base}$ performs slightly better with nSEBBs (0.393 vs. 0.390). These results suggest that our proposed nSEBBs efficiently preserves essential event boundaries without relying on computationally intensive operations.

However, both dual‐threshold variants (cSEBBs$_D$, nSEBBs$_D$) consistently underperform their 
single‐threshold counterparts. For MT-CRNN, PSDS$_1$ falls to 0.381 under cSEBBs$_D$ and to 0.371 under nSEBBs$_D$, compared with 0.389 under standard cSEBBs; Similar degradations occur for 
CRNN+WavLM$_\text{base}$ (0.360 and 0.344 vs. 0.371). Moreover, in the dual-model fusion as shown in 
Table \ref{data6}, nSEBBs again matches or slightly exceeds cSEBBs: CRNN+BEATs+WavLM$_{base}$ 
rises from PSDS$_1$ = 0.499 (cSEBBs) to 0.500 (nSEBBs), and CRNN+BEATs+WavLM$_{base+}$ from 0.492 to 
0.494. Here too, nSEBBs$_D$ and 
cSEBBs$_D$ lag behind. These results indicate that nSEBBs preserves essential event‐boundary refinement at a 
fraction of the computational cost, making it an attractive, efficiency‐driven choice for resource‐constrained 
SED deployments. 

\begin{table}[h]
    \centering
    \renewcommand{\arraystretch}{1.3}
    \caption{Performance comparison of post-processing methods in CRNNs integrated individual SSL representation.}
    \label{data5}
    \begin{tabular}{lccccc}
        \toprule
        \textbf{Model} & \textbf{baseline} & \textbf{cSEBBs} & \textbf{nSEBBs} & \textbf{cSEBBs$_D$}& \textbf{nSEBBs$_D$}\\
        \midrule
        MT-CRNN         & 0.353 & \textbf{0.389} & \textbf{0.389} & 0.381 & 0.371\\
        CRNN+BEATs      & 0.497 & \textbf{0.520} & 0.504 & 0.510 & 0.493\\
        CRNN+wav2vec2.0$_{\text{base}}$  & 0.350 & \textbf{0.390} & 0.384 & 0.377 & 0.372\\
        CRNN+HuBERT$_{\text{base}}$       & 0.352 & 0.390 & \textbf{0.393} & 0.386 & 0.365 \\
        CRNN+Dasheng$_{\text{base}}$      & 0.419 & \textbf{0.441} & 0.435 & 0.426 & 0.416\\
        CRNN+WavLM$_{\text{base}}$        & 0.342 & \textbf{0.373} & 0.371& 0.360 & 0.344\\
        CRNN+WavLM$_{\text{base+}}$      & 0.342 & \textbf{0.379} & 0.366 & 0.368 & 0.336\\
        \bottomrule
    \end{tabular}
\end{table}

\begin{table}[htbp]
    \centering
    \renewcommand{\arraystretch}{1.3}
    \caption{Performance comparison of post-processing methods in dual-model fusion framework.}
    \label{data6}
    \begin{tabular}{lccccc}
        \toprule
        \textbf{Model} & \textbf{baseline} & \textbf{cSEBBs} & \textbf{nSEBBs} & \textbf{cSEBBs$_D$}& \textbf{nSEBBs$_D$}\\
        \midrule
        CRNN+BEATs+wav2vec2.0$_{\text{base}}$& 0.471 & \textbf{0.494} & 0.489 & 0.476 & 0.468\\
        CRNN+BEATs+HuBERT$_{\text{base}}$& 0.475 & 0.493 & \textbf{0.495} & 0.492 & 0.481\\
        CRNN+BEATs+Dasheng$_{\text{base}}$& 0.452 & 0.467 & \textbf{0.469} & 0.463 & 0.443\\
        CRNN+BEATs+WavLM$_{\text{base}}$& 0.484 & 0.499 & \textbf{0.500} & 0.499 & 0.484\\
        CRNN+BEATs+WavLM$_{\text{base+}}$& 0.477 & 0.492 & 0.494& \textbf{0.495} & 0.481\\
        \bottomrule
    \end{tabular}
\end{table}

\section{Conclusion}\label{sec6}

This study provides a systematic framework for selecting and integrating SSL representation in sound event 
detection, providing industry practitioners with detailed guidance on leveraging SSL models to enhance SED 
system performance. Through extensive experiments, we demonstrate that dual-modal fusion (e.g., CRNN with both 
BEATs and WavLM embeddings) optimally balances complementary strengths, while CRNN+BEATs remains the top 
standalone architecture. To further enhance temporal localization, we propose the nSEBBs post-processing method, which dynamically adapts to event-specific characteristics and significantly outperforms traditional static thresholding approaches in both accuracy and efficiency. Our results also reveal that naïve full fusion all different SSL representations often degrades performance due to 
semantic misalignment between different embeddings and the resulting over-parameterization, 
underscoring the importance of strategic model pairing rather than indiscriminate aggregation. Overall, this work offers practical guidance for SSL model selection in SED by clarifying the trade-offs between computational efficiency and detection precision. Looking ahead, future research will explore event-aware fusion strategies and lightweight ensemble techniques to further optimize model integration. 


\bmhead{Data availability }
The authors declare that all training and testing data and codes supporting this study are available from the first author upon reasonable request. All other data supporting this study are available within the article.  

\section{Declarations}

\bmhead{Competing interest }
The authors declare no competing interests.

\bmhead{Funding}
The work is supported by the National Natural Science Foundation of China (Grant No.62071302).

\bmhead{Author contribution}
All authors contributed to the conception and design of the study. Hanfang Cui and Longfei Song performed material preparation, data collection, and analysis; Hanfang Cui and Li Li wrote the first draft of the manuscript;
Yanhua Long and Dongxing Xu reviewed and supervised this article. 

\noindent

\bibliography{sn-bibliography}


\begin{thebibliography}{41}
\ifx \bisbn   \undefined \def \bisbn  #1{ISBN #1}\fi
\ifx \binits  \undefined \def \binits#1{#1}\fi
\ifx \bauthor  \undefined \def \bauthor#1{#1}\fi
\ifx \batitle  \undefined \def \batitle#1{#1}\fi
\ifx \bjtitle  \undefined \def \bjtitle#1{#1}\fi
\ifx \bvolume  \undefined \def \bvolume#1{\textbf{#1}}\fi
\ifx \byear  \undefined \def \byear#1{#1}\fi
\ifx \bissue  \undefined \def \bissue#1{#1}\fi
\ifx \bfpage  \undefined \def \bfpage#1{#1}\fi
\ifx \blpage  \undefined \def \blpage #1{#1}\fi
\ifx \burl  \undefined \def \burl#1{\textsf{#1}}\fi
\ifx \doiurl  \undefined \def \doiurl#1{\url{https://doi.org/#1}}\fi
\ifx \betal  \undefined \def \betal{\textit{et al.}}\fi
\ifx \binstitute  \undefined \def \binstitute#1{#1}\fi
\ifx \binstitutionaled  \undefined \def \binstitutionaled#1{#1}\fi
\ifx \bctitle  \undefined \def \bctitle#1{#1}\fi
\ifx \beditor  \undefined \def \beditor#1{#1}\fi
\ifx \bpublisher  \undefined \def \bpublisher#1{#1}\fi
\ifx \bbtitle  \undefined \def \bbtitle#1{#1}\fi
\ifx \bedition  \undefined \def \bedition#1{#1}\fi
\ifx \bseriesno  \undefined \def \bseriesno#1{#1}\fi
\ifx \blocation  \undefined \def \blocation#1{#1}\fi
\ifx \bsertitle  \undefined \def \bsertitle#1{#1}\fi
\ifx \bsnm \undefined \def \bsnm#1{#1}\fi
\ifx \bsuffix \undefined \def \bsuffix#1{#1}\fi
\ifx \bparticle \undefined \def \bparticle#1{#1}\fi
\ifx \barticle \undefined \def \barticle#1{#1}\fi
\bibcommenthead
\ifx \bconfdate \undefined \def \bconfdate #1{#1}\fi
\ifx \botherref \undefined \def \botherref #1{#1}\fi
\ifx \url \undefined \def \url#1{\textsf{#1}}\fi
\ifx \bchapter \undefined \def \bchapter#1{#1}\fi
\ifx \bbook \undefined \def \bbook#1{#1}\fi
\ifx \bcomment \undefined \def \bcomment#1{#1}\fi
\ifx \oauthor \undefined \def \oauthor#1{#1}\fi
\ifx \citeauthoryear \undefined \def \citeauthoryear#1{#1}\fi
\ifx \endbibitem  \undefined \def \endbibitem {}\fi
\ifx \bconflocation  \undefined \def \bconflocation#1{#1}\fi
\ifx \arxivurl  \undefined \def \arxivurl#1{\textsf{#1}}\fi
\csname PreBibitemsHook\endcsname

\bibitem[\protect\citeauthoryear{Ashraf et~al.}{2015}]{ashraf2015audio}
\begin{bchapter}
\bauthor{\bsnm{Ashraf}, \binits{K.}},
\bauthor{\bsnm{Elizalde}, \binits{B.}},
\bauthor{\bsnm{Iandola}, \binits{F.}},
\bauthor{\bsnm{Moskewicz}, \binits{M.}},
\bauthor{\bsnm{Bernd}, \binits{J.}},
\bauthor{\bsnm{Friedland}, \binits{G.}},
\bauthor{\bsnm{Keutzer}, \binits{K.}}:
\bctitle{Audio-based multimedia event detection with dnns sparse sampling}.
In: \bbtitle{Proceedings of the 5th ACM International Conference on Multimedia Retrieval},
pp. \bfpage{611}--\blpage{614}
(\byear{2015}).
\doiurl{10.1145/2671188.2749396}
\end{bchapter}
\endbibitem

\bibitem[\protect\citeauthoryear{Bilen et~al.}{2020}]{bilen2020framework}
\begin{bchapter}
\bauthor{\bsnm{Bilen}, \binits{{\c{C}}.}},
\bauthor{\bsnm{Ferroni}, \binits{G.}},
\bauthor{\bsnm{Tuveri}, \binits{F.}},
\bauthor{\bsnm{Azcarreta}, \binits{J.}},
\bauthor{\bsnm{Krstulovi{\'c}}, \binits{S.}}:
\bctitle{A framework for the robust evaluation of sound event detection}.
In: \bbtitle{ICASSP 2020 - IEEE International Conference on Acoustics, Speech and Signal Processing (ICASSP)},
pp. \bfpage{61}--\blpage{65}
(\byear{2020}).
\doiurl{10.1109/ICASSP40776.2020.9052995}
\end{bchapter}
\endbibitem

\bibitem[\protect\citeauthoryear{Bello et~al.}{2018}]{bello2018sound}
\begin{bchapter}
\bauthor{\bsnm{Bello}, \binits{J.P.}},
\bauthor{\bsnm{Mydlarz}, \binits{C.}},
\bauthor{\bsnm{Salamon}, \binits{J.}}:
\bctitle{Sound analysis in smart cities}.
In: \beditor{\bsnm{Virtanen}, \binits{T.}},
\beditor{\bsnm{Plumbley}, \binits{M.D.}},
\beditor{\bsnm{Ellis}, \binits{D.}} (eds.)
\bbtitle{Computational Analysis of Sound Scenes and Events},
pp. \bfpage{373}--\blpage{397}.
\bpublisher{Springer},
\blocation{Cham}
(\byear{2018})
\end{bchapter}
\endbibitem

\bibitem[\protect\citeauthoryear{Bello et~al.}{2019}]{bello2019sonyc}
\begin{barticle}
\bauthor{\bsnm{Bello}, \binits{J.P.}},
\bauthor{\bsnm{Silva}, \binits{C.}},
\bauthor{\bsnm{Nov}, \binits{O.}},
\bauthor{\bsnm{Dubois}, \binits{R.L.}},
\bauthor{\bsnm{Arora}, \binits{A.}},
\bauthor{\bsnm{Salamon}, \binits{J.}},
\bauthor{\bsnm{Mydlarz}, \binits{C.}},
\bauthor{\bsnm{Doraiswamy}, \binits{H.}}:
\batitle{Sonyc: A system for monitoring, analyzing, and mitigating urban noise pollution}.
\bjtitle{Communications of the ACM}
\bvolume{62},
\bfpage{68}--\blpage{77}
(\byear{2019})
\doiurl{10.1145/3224204}
\end{barticle}
\endbibitem

\bibitem[\protect\citeauthoryear{Baevski et~al.}{2020}]{baevski2020wav2vec}
\begin{bchapter}
\bauthor{\bsnm{Baevski}, \binits{A.}},
\bauthor{\bsnm{Zhou}, \binits{Y.}},
\bauthor{\bsnm{Mohamed}, \binits{A.}},
\bauthor{\bsnm{Auli}, \binits{M.}}:
\bctitle{wav2vec 2.0: A framework for self-supervised learning of speech representations}.
In: \bbtitle{Advances in Neural Information Processing Systems},
vol. \bseriesno{33},
pp. \bfpage{12449}--\blpage{12460}
(\byear{2020})
\end{bchapter}
\endbibitem

\bibitem[\protect\citeauthoryear{Clarkson et~al.}{1998}]{clarkson1998auditory}
\begin{barticle}
\bauthor{\bsnm{Clarkson}, \binits{B.}},
\bauthor{\bsnm{Pentl}, \binits{A.}},
\bauthor{\bsnm{Sawhney}, \binits{N.}}:
\batitle{Auditory context awareness via wearable computing}.
\bjtitle{Energy}
\bvolume{400},
\bfpage{20}
(\byear{1998})
\end{barticle}
\endbibitem

\bibitem[\protect\citeauthoryear{Cai et~al.}{2024}]{cai2024mat}
\begin{bchapter}
\bauthor{\bsnm{Cai}, \binits{P.}},
\bauthor{\bsnm{Song}, \binits{Y.}},
\bauthor{\bsnm{Li}, \binits{K.}},
\bauthor{\bsnm{Song}, \binits{H.}},
\bauthor{\bsnm{McLoughlin}, \binits{I.}}:
\bctitle{{MAT-SED}: A masked audio transformer with masked-reconstruction based pre-training for sound event detection}.
In: \bbtitle{Proceedings of Interspeech 2024},
pp. \bfpage{557}--\blpage{561}
(\byear{2024}).
\doiurl{10.21437/Interspeech.2024-714}
\end{bchapter}
\endbibitem

\bibitem[\protect\citeauthoryear{Chen et~al.}{2022}]{chen2022wavlm}
\begin{barticle}
\bauthor{\bsnm{Chen}, \binits{S.}},
\bauthor{\bsnm{Wang}, \binits{C.}},
\bauthor{\bsnm{Chen}, \binits{Z.}},
\bauthor{\bsnm{Wu}, \binits{Y.}},
\bauthor{\bsnm{Liu}, \binits{S.}},
\bauthor{\bsnm{Chen}, \binits{Z.}},
\bauthor{\bsnm{Li}, \binits{J.}},
\bauthor{\bsnm{Kanda}, \binits{N.}},
\bauthor{\bsnm{Yoshioka}, \binits{T.}},
\bauthor{\bsnm{Xiao}, \binits{X.}},
\bauthor{\bsnm{Wu}, \binits{J.}},
\bauthor{\bsnm{Zhou}, \binits{L.}},
\bauthor{\bsnm{Ren}, \binits{S.}},
\bauthor{\bsnm{Qian}, \binits{Y.}},
\bauthor{\bsnm{Qian}, \binits{Y.}},
\bauthor{\bsnm{Wu}, \binits{J.}},
\bauthor{\bsnm{Zeng}, \binits{M.}},
\bauthor{\bsnm{Yu}, \binits{X.}},
\bauthor{\bsnm{Wei}, \binits{F.}}:
\batitle{{WavLM}: Large-scale self-supervised pre-training for full stack speech processing}.
\bjtitle{IEEE Journal of Selected Topics in Signal Processing}
\bvolume{16},
\bfpage{1505}--\blpage{1518}
(\byear{2022})
\doiurl{10.1109/JSTSP.2022.3188113}
\end{barticle}
\endbibitem

\bibitem[\protect\citeauthoryear{Chen et~al.}{2022}]{chen2022beats}
\begin{botherref}
\oauthor{\bsnm{Chen}, \binits{S.}},
\oauthor{\bsnm{Wu}, \binits{Y.}},
\oauthor{\bsnm{Wang}, \binits{C.}},
\oauthor{\bsnm{Liu}, \binits{S.}},
\oauthor{\bsnm{Tompkins}, \binits{D.}},
\oauthor{\bsnm{Chen}, \binits{Z.}},
\oauthor{\bsnm{Che}, \binits{W.}},
\oauthor{\bsnm{Yu}, \binits{X.}},
\oauthor{\bsnm{Wei}, \binits{F.}}:
{BEATs}: Audio pre-training with acoustic tokenizers.
arXiv preprint arXiv:2212.09058
(2022)
\end{botherref}
\endbibitem

\bibitem[\protect\citeauthoryear{{DCASE Community}}{2023}]{dcase2023}
\begin{botherref}
\oauthor{\bsnm{{DCASE Community}}}:
{DCASE Event Detection with Weak Soundscapes}.
\url{https://dcase.community/challenge2023/task-sound-event-detection-with-weak-labels-and-synthetic-soundscapes}.
Accessed 12 Oct 2024
(2023)
\end{botherref}
\endbibitem

\bibitem[\protect\citeauthoryear{Dohi et~al.}{2023}]{dohi2023description}
\begin{botherref}
\oauthor{\bsnm{Dohi}, \binits{K.}},
\oauthor{\bsnm{Imoto}, \binits{K.}},
\oauthor{\bsnm{Harada}, \binits{N.}},
\oauthor{\bsnm{Niizumi}, \binits{D.}},
\oauthor{\bsnm{Koizumi}, \binits{Y.}},
\oauthor{\bsnm{Nishida}, \binits{T.}},
\oauthor{\bsnm{Purohit}, \binits{H.}},
\oauthor{\bsnm{Tanabe}, \binits{R.}},
\oauthor{\bsnm{Endo}, \binits{T.}},
\oauthor{\bsnm{Kawaguchi}, \binits{Y.}}:
Description and discussion on {DCASE 2023 Challenge Task 2}: First-shot unsupervised anomalous sound detection for machine condition monitoring.
arXiv preprint arXiv:2305.07828
(2023)
\end{botherref}
\endbibitem

\bibitem[\protect\citeauthoryear{Debes et~al.}{2016}]{debes2016monitoring}
\begin{barticle}
\bauthor{\bsnm{Debes}, \binits{C.}},
\bauthor{\bsnm{Merentitis}, \binits{A.}},
\bauthor{\bsnm{Sukhanov}, \binits{S.}},
\bauthor{\bsnm{Niessen}, \binits{M.}},
\bauthor{\bsnm{Frangiadakis}, \binits{N.}},
\bauthor{\bsnm{Bauer}, \binits{A.}}:
\batitle{Monitoring activities of daily living in smart homes: Understanding human behavior}.
\bjtitle{IEEE Signal Processing Magazine}
\bvolume{33},
\bfpage{81}--\blpage{94}
(\byear{2016})
\doiurl{10.1109/MSP.2015.2503881}
\end{barticle}
\endbibitem

\bibitem[\protect\citeauthoryear{Dinkel et~al.}{2024}]{dinkel2024dasheng}
\begin{bchapter}
\bauthor{\bsnm{Dinkel}, \binits{H.}},
\bauthor{\bsnm{Yan}, \binits{Z.}},
\bauthor{\bsnm{Wang}, \binits{Y.}},
\bauthor{\bsnm{Zhang}, \binits{J.}},
\bauthor{\bsnm{Wang}, \binits{Y.}},
\bauthor{\bsnm{Wang}, \binits{B.}}:
\bctitle{Scaling up masked audio encoder learning for general audio classification}.
In: \bbtitle{Proceedings of Interspeech 2024},
pp. \bfpage{547}--\blpage{551}
(\byear{2024}).
\doiurl{10.21437/Interspeech.2024-246}
\end{bchapter}
\endbibitem

\bibitem[\protect\citeauthoryear{Ebbers et~al.}{2024}]{ebbers2024sound}
\begin{bchapter}
\bauthor{\bsnm{Ebbers}, \binits{J.}},
\bauthor{\bsnm{Germain}, \binits{F.G.}},
\bauthor{\bsnm{Wichern}, \binits{G.}},
\bauthor{\bsnm{Le~Roux}, \binits{J.}}:
\bctitle{Sound event bounding boxes}.
In: \bbtitle{Proceedings of Interspeech 2024},
pp. \bfpage{562}--\blpage{566}
(\byear{2024}).
\doiurl{10.21437/Interspeech.2024-2075}
\end{bchapter}
\endbibitem

\bibitem[\protect\citeauthoryear{Ebbers and Haeb-Umbach}{2022}]{ebbers2022pre}
\begin{botherref}
\oauthor{\bsnm{Ebbers}, \binits{J.}},
\oauthor{\bsnm{Haeb-Umbach}, \binits{R.}}:
Pre-training and self-training for sound event detection in domestic environments.
Technical report,
DCASE2022 Challenge
(2022)
\end{botherref}
\endbibitem

\bibitem[\protect\citeauthoryear{Gong et~al.}{2021}]{gong2021ast}
\begin{bchapter}
\bauthor{\bsnm{Gong}, \binits{Y.}},
\bauthor{\bsnm{Chung}, \binits{Y.-A.}},
\bauthor{\bsnm{Glass}, \binits{J.}}:
\bctitle{{AST}: Audio spectrogram transformer}.
In: \bbtitle{Proceedings of Interspeech 2021},
pp. \bfpage{571}--\blpage{575}
(\byear{2021}).
\doiurl{10.21437/Interspeech.2021-698}
\end{bchapter}
\endbibitem

\bibitem[\protect\citeauthoryear{Hsu et~al.}{2021}]{hsu2021hubert}
\begin{barticle}
\bauthor{\bsnm{Hsu}, \binits{W.-N.}},
\bauthor{\bsnm{Bolte}, \binits{B.}},
\bauthor{\bsnm{Tsai}, \binits{Y.-H.H.}},
\bauthor{\bsnm{Lakhotia}, \binits{K.}},
\bauthor{\bsnm{Salakhutdinov}, \binits{R.}},
\bauthor{\bsnm{Mohamed}, \binits{A.}}:
\batitle{{HuBERT}: Self-supervised speech representation learning by masked prediction of hidden units}.
\bjtitle{IEEE/ACM Transactions on Audio, Speech, and Language Processing}
\bvolume{29},
\bfpage{3451}--\blpage{3460}
(\byear{2021})
\doiurl{10.1109/TASLP.2021.3122291}
\end{barticle}
\endbibitem

\bibitem[\protect\citeauthoryear{He et~al.}{2022}]{he2022masked}
\begin{bchapter}
\bauthor{\bsnm{He}, \binits{K.}},
\bauthor{\bsnm{Chen}, \binits{X.}},
\bauthor{\bsnm{Xie}, \binits{S.}},
\bauthor{\bsnm{Li}, \binits{Y.}},
\bauthor{\bsnm{Doll{\'a}r}, \binits{P.}},
\bauthor{\bsnm{Girshick}, \binits{R.}}:
\bctitle{Masked autoencoders are scalable vision learners}.
In: \bbtitle{Proceedings of the IEEE/CVF Conference on Computer Vision and Pattern Recognition},
pp. \bfpage{16000}--\blpage{16009}
(\byear{2022}).
\doiurl{10.1109/CVPR52688.2022.01553}
\end{bchapter}
\endbibitem

\bibitem[\protect\citeauthoryear{Kong et~al.}{2020}]{kong2020panns}
\begin{barticle}
\bauthor{\bsnm{Kong}, \binits{Q.}},
\bauthor{\bsnm{Cao}, \binits{Y.}},
\bauthor{\bsnm{Iqbal}, \binits{T.}},
\bauthor{\bsnm{Wang}, \binits{Y.}},
\bauthor{\bsnm{Wang}, \binits{W.}},
\bauthor{\bsnm{Plumbley}, \binits{M.D.}}:
\batitle{{PANNs}: Large-scale pretrained audio neural networks for audio pattern recognition}.
\bjtitle{IEEE/ACM Transactions on Audio, Speech, and Language Processing}
\bvolume{28},
\bfpage{2880}--\blpage{2894}
(\byear{2020})
\doiurl{10.1109/TASLP.2020.3030497}
\end{barticle}
\endbibitem

\bibitem[\protect\citeauthoryear{Khandelwal et~al.}{2024}]{khelwal2024sound}
\begin{barticle}
\bauthor{\bsnm{Khandelwal}, \binits{T.}},
\bauthor{\bsnm{Das}, \binits{R.K.}},
\bauthor{\bsnm{Chng}, \binits{E.S.}}:
\batitle{Sound event detection: A journey through dcase challenge series}.
\bjtitle{APSIPA Transactions on Signal and Information Processing}
\bvolume{13},
\bfpage{51}
(\byear{2024})
\doiurl{10.1561/116.00000051}
\end{barticle}
\endbibitem

\bibitem[\protect\citeauthoryear{Kim et~al.}{2023}]{kim2023semi}
\begin{botherref}
\oauthor{\bsnm{Kim}, \binits{J.W.}},
\oauthor{\bsnm{Son}, \binits{S.W.}},
\oauthor{\bsnm{Song}, \binits{Y.}},
\oauthor{\bsnm{Kim}, \binits{H.K.}},
\oauthor{\bsnm{Song}, \binits{I.H.}},
\oauthor{\bsnm{Lim}, \binits{J.E.}}:
Semi-supervised learning-based sound event detection using frequency dynamic convolution with large kernel attention for {DCASE Challenge 2023 Task 4}.
arXiv preprint arXiv:2306.06461
(2023)
\end{botherref}
\endbibitem

\bibitem[\protect\citeauthoryear{LeCun et~al.}{1998}]{lecun1998gradient}
\begin{barticle}
\bauthor{\bsnm{LeCun}, \binits{Y.}},
\bauthor{\bsnm{Bottou}, \binits{L.}},
\bauthor{\bsnm{Bengio}, \binits{Y.}},
\bauthor{\bsnm{Haffner}, \binits{P.}}:
\batitle{Gradient-based learning applied to document recognition}.
\bjtitle{Proceedings of the IEEE}
\bvolume{86},
\bfpage{2278}--\blpage{2324}
(\byear{1998})
\doiurl{10.1109/5.726791}
\end{barticle}
\endbibitem

\bibitem[\protect\citeauthoryear{Lv et~al.}{2023}]{lv2023unsupervised}
\begin{botherref}
\oauthor{\bsnm{Lv}, \binits{Z.}},
\oauthor{\bsnm{Han}, \binits{B.}},
\oauthor{\bsnm{Chen}, \binits{Z.}},
\oauthor{\bsnm{Qian}, \binits{Y.}},
\oauthor{\bsnm{Ding}, \binits{J.}},
\oauthor{\bsnm{Liu}, \binits{J.}}:
Unsupervised anomalous detection based on unsupervised pretrained models.
Technical report,
DCASE2023 Challenge
(2023)
\end{botherref}
\endbibitem

\bibitem[\protect\citeauthoryear{Lawrance and Lewis}{1977}]{lawrance1977exponential}
\begin{barticle}
\bauthor{\bsnm{Lawrance}, \binits{A.J.}},
\bauthor{\bsnm{Lewis}, \binits{P.A.W.}}:
\batitle{An exponential moving-average sequence and point process (ema1)}.
\bjtitle{Journal of Applied Probability}
\bvolume{14},
\bfpage{98}--\blpage{113}
(\byear{1977})
\doiurl{10.2307/3213263}
\end{barticle}
\endbibitem

\bibitem[\protect\citeauthoryear{Liang et~al.}{2022}]{liang2022joint}
\begin{barticle}
\bauthor{\bsnm{Liang}, \binits{Y.}},
\bauthor{\bsnm{Long}, \binits{Y.}},
\bauthor{\bsnm{Li}, \binits{Y.}},
\bauthor{\bsnm{Liang}, \binits{J.}},
\bauthor{\bsnm{Wang}, \binits{Y.}}:
\batitle{Joint framework with deep feature distillation and adaptive focal loss for weakly supervised audio tagging and acoustic event detection}.
\bjtitle{Digital Signal Processing}
\bvolume{123},
\bfpage{103446}
(\byear{2022})
\doiurl{10.1016/j.dsp.2022.103446}
\end{barticle}
\endbibitem

\bibitem[\protect\citeauthoryear{Li et~al.}{2024}]{li2024self}
\begin{barticle}
\bauthor{\bsnm{Li}, \binits{X.}},
\bauthor{\bsnm{Shao}, \binits{N.}},
\bauthor{\bsnm{Li}, \binits{X.}}:
\batitle{Self-supervised audio teacher-student transformer for both clip-level and frame-level tasks}.
\bjtitle{IEEE/ACM Transactions on Audio, Speech, and Language Processing}
\bvolume{32},
\bfpage{1336}--\blpage{1351}
(\byear{2024})
\doiurl{10.1109/TASLP.2024.3352248}
\end{barticle}
\endbibitem

\bibitem[\protect\citeauthoryear{Liu et~al.}{2024}]{liu2024icsd}
\begin{botherref}
\oauthor{\bsnm{Liu}, \binits{Q.}},
\oauthor{\bsnm{Song}, \binits{L.}},
\oauthor{\bsnm{Xu}, \binits{D.}},
\oauthor{\bsnm{Long}, \binits{Y.}}:
{ICSD}: An open-source dataset for infant cry and snoring detection.
arXiv preprint arXiv:2408.10561
(2024)
\end{botherref}
\endbibitem

\bibitem[\protect\citeauthoryear{Nam et~al.}{2022}]{nam2022frequency}
\begin{bchapter}
\bauthor{\bsnm{Nam}, \binits{H.}},
\bauthor{\bsnm{Kim}, \binits{S.-H.}},
\bauthor{\bsnm{Ko}, \binits{B.-Y.}},
\bauthor{\bsnm{Park}, \binits{Y.-H.}}:
\bctitle{Frequency dynamic convolution: Frequency-adaptive pattern recognition for sound event detection}.
In: \bbtitle{Proceedings of Interspeech 2022},
pp. \bfpage{2763}--\blpage{2767}
(\byear{2022}).
\doiurl{10.21437/Interspeech.2022-10127}
\end{bchapter}
\endbibitem

\bibitem[\protect\citeauthoryear{Schneider et~al.}{2019}]{schneider2019wav2vec}
\begin{bchapter}
\bauthor{\bsnm{Schneider}, \binits{S.}},
\bauthor{\bsnm{Baevski}, \binits{A.}},
\bauthor{\bsnm{Collobert}, \binits{R.}},
\bauthor{\bsnm{Auli}, \binits{M.}}:
\bctitle{wav2vec: Unsupervised pre-training for speech recognition}.
In: \bbtitle{Proceedings of Interspeech 2019},
pp. \bfpage{3465}--\blpage{3469}
(\byear{2019}).
\doiurl{10.21437/Interspeech.2019-1873}
\end{bchapter}
\endbibitem

\bibitem[\protect\citeauthoryear{Shi et~al.}{2016}]{shi2016end}
\begin{barticle}
\bauthor{\bsnm{Shi}, \binits{B.}},
\bauthor{\bsnm{Bai}, \binits{X.}},
\bauthor{\bsnm{Yao}, \binits{C.}}:
\batitle{An end-to-end trainable neural network for image-based sequence recognition and its application to scene text recognition}.
\bjtitle{IEEE Transactions on Pattern Analysis and Machine Intelligence}
\bvolume{39},
\bfpage{2298}--\blpage{2304}
(\byear{2016})
\doiurl{10.1109/TPAMI.2016.2646371}
\end{barticle}
\endbibitem

\bibitem[\protect\citeauthoryear{Singh et~al.}{2021}]{singh2021polyphonic}
\begin{bchapter}
\bauthor{\bsnm{Singh}, \binits{U.}},
\bauthor{\bsnm{Dash}, \binits{D.D.}},
\bauthor{\bsnm{Sharma}, \binits{M.}},
\bauthor{\bsnm{Mishra}, \binits{S.}},
\bauthor{\bsnm{Malarvizhi}, \binits{S.}},
\bauthor{\bsnm{Tiwari}, \binits{S.}},
\bauthor{\bsnm{Shankarappa}, \binits{R.T.}}:
\bctitle{Polyphonic sound event detection and classification using convolutional recurrent neural network with mean teacher}.
In: \bbtitle{2021 12th International Conference on Computing Communication and Networking Technologies (ICCCNT)},
pp. \bfpage{1}--\blpage{4}
(\byear{2021}).
\doiurl{10.1109/ICCCNT51525.2021.9579677}
\end{bchapter}
\endbibitem

\bibitem[\protect\citeauthoryear{Shao et~al.}{2024}]{shao2024fine}
\begin{bchapter}
\bauthor{\bsnm{Shao}, \binits{N.}},
\bauthor{\bsnm{Li}, \binits{X.}},
\bauthor{\bsnm{Li}, \binits{X.}}:
\bctitle{Fine-tune the pretrained atst model for sound event detection}.
In: \bbtitle{ICASSP 2024 - IEEE International Conference on Acoustics, Speech and Signal Processing (ICASSP)},
pp. \bfpage{911}--\blpage{915}
(\byear{2024}).
\doiurl{10.1109/ICASSP48485.2024.10446159}
\end{bchapter}
\endbibitem

\bibitem[\protect\citeauthoryear{Turian et~al.}{2022}]{turian2022hear}
\begin{bchapter}
\bauthor{\bsnm{Turian}, \binits{J.}},
\bauthor{\bsnm{Shier}, \binits{J.}},
\bauthor{\bsnm{Khan}, \binits{H.R.}},
\bauthor{\bsnm{Raj}, \binits{B.}},
\bauthor{\bsnm{Schuller}, \binits{B.W.}},
\bauthor{\bsnm{Steinmetz}, \binits{C.J.}},
\bauthor{\bsnm{Malloy}, \binits{C.}},
\bauthor{\bsnm{Tzanetakis}, \binits{G.}},
\bauthor{\bsnm{Velarde}, \binits{G.}},
\bauthor{\bsnm{McNally}, \binits{K.}}, \betal:
\bctitle{{HEAR}: Holistic evaluation of audio representations}.
In: \bbtitle{NeurIPS 2021 Competitions and Demonstrations Track},
pp. \bfpage{125}--\blpage{145}
(\byear{2022}).
\bcomment{PMLR}
\end{bchapter}
\endbibitem

\bibitem[\protect\citeauthoryear{Turpault et~al.}{2019}]{turpault2019sound}
\begin{bchapter}
\bauthor{\bsnm{Turpault}, \binits{N.}},
\bauthor{\bsnm{Serizel}, \binits{R.}},
\bauthor{\bsnm{Shah}, \binits{A.P.}},
\bauthor{\bsnm{Salamon}, \binits{J.}}:
\bctitle{Sound event detection in domestic environments with weakly labeled data and soundscape synthesis}.
In: \bbtitle{Workshop on Detection and Classification of Acoustic Scenes and Events (DCASE)},
pp. \bfpage{253}--\blpage{257}
(\byear{2019}).
\doiurl{10.33682/006b-jx26}
\end{bchapter}
\endbibitem

\bibitem[\protect\citeauthoryear{Tarvainen and Valpola}{2017}]{tarvainen2017mean}
\begin{bchapter}
\bauthor{\bsnm{Tarvainen}, \binits{A.}},
\bauthor{\bsnm{Valpola}, \binits{H.}}:
\bctitle{Mean teachers are better role models: Weight-averaged consistency targets improve semi-supervised deep learning results}.
In: \bbtitle{Advances in Neural Information Processing Systems},
vol. \bseriesno{30}
(\byear{2017})
\end{bchapter}
\endbibitem

\bibitem[\protect\citeauthoryear{van~den Oord et~al.}{2018}]{oord2018representation}
\begin{barticle}
\bauthor{\bsnm{Oord}, \binits{A.}},
\bauthor{\bsnm{Li}, \binits{Y.}},
\bauthor{\bsnm{Vinyals}, \binits{O.}}:
\batitle{Representation learning with contrastive predictive coding}.
\bjtitle{arXiv preprint arXiv:1807.03748}
(\byear{2018})
\doiurl{10.48550/arXiv.1807.03748}
\end{barticle}
\endbibitem

\bibitem[\protect\citeauthoryear{Vaswani et~al.}{2017}]{vaswani2017attention}
\begin{bchapter}
\bauthor{\bsnm{Vaswani}, \binits{A.}},
\bauthor{\bsnm{Shazeer}, \binits{N.}},
\bauthor{\bsnm{Parmar}, \binits{N.}},
\bauthor{\bsnm{Uszkoreit}, \binits{J.}},
\bauthor{\bsnm{Jones}, \binits{L.}},
\bauthor{\bsnm{Gomez}, \binits{A.N.}},
\bauthor{\bsnm{Kaiser}, \binits{{\L}.}},
\bauthor{\bsnm{Polosukhin}, \binits{I.}}:
\bctitle{Attention is all you need}.
In: \bbtitle{Advances in Neural Information Processing Systems},
vol. \bseriesno{30},
pp. \bfpage{5998}--\blpage{6008}
(\byear{2017})
\end{bchapter}
\endbibitem

\bibitem[\protect\citeauthoryear{Wang et~al.}{2024}]{wang2024cross}
\begin{bchapter}
\bauthor{\bsnm{Wang}, \binits{Y.}},
\bauthor{\bsnm{Zheng}, \binits{H.}},
\bauthor{\bsnm{Sun}, \binits{Q.}},
\bauthor{\bsnm{Ma}, \binits{Y.}},
\bauthor{\bsnm{Zhu}, \binits{S.}},
\bauthor{\bsnm{Zhang}, \binits{L.}},
\bauthor{\bsnm{Zhang}, \binits{W.-Q.}}:
\bctitle{Cross-lingual alzheimer's disease detection based on scale criteria}.
In: \bbtitle{2024 IEEE 14th International Symposium on Chinese Spoken Language Processing (ISCSLP)},
pp. \bfpage{491}--\blpage{495}
(\byear{2024}).
\doiurl{10.1109/ISCSLP63861.2024.10800047}
\end{bchapter}
\endbibitem

\bibitem[\protect\citeauthoryear{Zhao et~al.}{2024}]{zhao2024sound}
\begin{barticle}
\bauthor{\bsnm{Zhao}, \binits{D.}},
\bauthor{\bsnm{Ding}, \binits{K.}},
\bauthor{\bsnm{Qi}, \binits{X.}},
\bauthor{\bsnm{Chen}, \binits{Y.}},
\bauthor{\bsnm{Feng}, \binits{H.}}:
\batitle{Sound event localization and detection based on deep learning}.
\bjtitle{Journal of Systems Engineering and Electronics}
\bvolume{35},
\bfpage{294}--\blpage{301}
(\byear{2024})
\doiurl{10.23919/JSEE.2023.000110}
\end{barticle}
\endbibitem

\bibitem[\protect\citeauthoryear{Zheng et~al.}{2021}]{zheng2021effective}
\begin{bchapter}
\bauthor{\bsnm{Zheng}, \binits{X.}},
\bauthor{\bsnm{Song}, \binits{Y.}},
\bauthor{\bsnm{Dai}, \binits{L.-R.}},
\bauthor{\bsnm{McLoughlin}, \binits{I.}},
\bauthor{\bsnm{Liu}, \binits{L.}}:
\bctitle{An effective mutual mean teaching based domain adaptation method for sound event detection}.
In: \bbtitle{Proceedings of Interspeech 2021},
pp. \bfpage{556}--\blpage{560}
(\byear{2021}).
\doiurl{10.21437/Interspeech.2021-281}
\end{bchapter}
\endbibitem

\bibitem[\protect\citeauthoryear{Zheng et~al.}{2025}]{zheng2025asit}
\begin{botherref}
\oauthor{\bsnm{Zheng}, \binits{Y.}},
\oauthor{\bsnm{Zhang}, \binits{R.}},
\oauthor{\bsnm{Atito}, \binits{S.}},
\oauthor{\bsnm{Yang}, \binits{S.}},
\oauthor{\bsnm{Wang}, \binits{W.}},
\oauthor{\bsnm{Mei}, \binits{Y.}}:
{ASiT-CRNN}: A method for sound event detection with fine-tuning of self-supervised pre-trained asit-based model.
Digital Signal Processing,
105055
(2025)
\doiurl{10.1016/j.dsp.2025.105055}
\end{botherref}
\endbibitem

\end{thebibliography}

\end{document}